\documentclass[a4paper,fleqn,usenatbib]{mnras}

\usepackage[T1]{fontenc}
\usepackage{ae,aecompl}

\usepackage{graphicx}	
\usepackage{amsmath}	
\usepackage{amssymb}	
\usepackage{subfigure}
\usepackage{csquotes}

\usepackage{longtable}
\usepackage{tabularx}
\usepackage{caption, booktabs}
\usepackage{setspace}
\usepackage{threeparttable}
\usepackage{threeparttablex}
\usepackage[utf8]{inputenc}
\usepackage{color}

\title[Metallicity of Supernova HII regions]{Metallicity Estimation of Core-Collapse Supernova HII Regions in Galaxies within 30 Mpc
} 

\author[R. Ganss et al.]
{R. Ganss,$^{1}$
J. L. Pledger,$^{1}$\thanks{E-mail: JPledger@uclan.ac.uk}
A. E. Sansom,$^{1}$
P. A. James,$^{2}$
J. Puls$^{3}$ 
\and and S. M. Habergham-Mawson$^{2}$
\\
$^{1}$Jeremiah Horrocks Institute, University of Central Lancashire, Preston, PR1 2HE, UK \\ 
$^{2}$Astrophysics Research Institute, Liverpool John Moores University, 146 Brownlow Hill, Liverpool, L3 5RF, UK \\
$^{3}$LMU München, Universitätssternwarte, Scheinerstr. 1, D-81679 München, Germany \\
}

\date{Accepted 2022 March 03. Received 2022 March 01; in original form 2022 January 14}

\pubyear{2022}

\begin{document}
\label{firstpage}
\pagerange{\pageref{firstpage}--\pageref{lastpage}}
\maketitle

\begin{abstract}
This work presents measurements of the local HII environment metallicities of core-collapse supernovae (SNe) within a luminosity distance of 30\,Mpc. 76 targets were observed at the Isaac Newton Telescope and environment metallicities could be measured for 65 targets using the N2 and O3N2 strong emission line method. The cumulative distribution functions (CDFs) of the environment metallicities of Type Ib and Ic SNe tend to higher metallicity than Type IIP, however Type Ic are also present at lower metallicities whereas Type Ib are not. The Type Ib frequency distribution  is narrower (standard deviation $\sim$0.06 dex) than the Ic and IIP distributions ($\sim$0.15 dex) giving some evidence for a significant fraction of single massive progenitor stars; the low metallicity of Type Ic suggests a significant fraction of compact binary progenitors. However, both the Kolmogorov-Smirnov test and the Anderson-Darling test indicate no statistical significance for a difference in the local metallicities of the three SN types. Monte-Carlo simulations reveal a strong sensitivity of these tests to the uncertainties of the derived metallicities. Given the uncertainties of the strong emission methods, the applicability of the tests seems limited. We extended our analysis with the data of the Type Ib/Ic/IIP SN sample from Galbany et al. (2018). The CDFs created with their sample confirm our CDFs very well. The statistical tests, combining our sample and the Galbany et al. (2018) sample, indicate a significant difference between Type Ib and Type IIP with <5\% probability that they are drawn from the same parent population.
\end{abstract}



\section{Introduction}

Massive stars are the key players in the chemical enrichment of the universe. Their strong stellar winds and their ultimate death as core-collapse supernovae (SNe) enrich the interstellar medium with chemical elements burned during a star’s lifetime and with new elements created in the SN explosion. To understand the details of the SN explosion process and the synthesis of heavy elements beyond iron, knowledge of the nature of SN progenitor stars is essential. It is commonly accepted that massive stars with masses >8 M$_{\sun}$ end their life in a core-collapse SN explosion (e.g. reviews by \citealt{crowther2007, smartt2009}), however the detailed links between the progenitor star parameters and the observed diversity of SN explosions are still uncertain.

Historically, the classification of SNe is mainly based on spectral features (e.g. \citealt{filippenko1997, turatto2003}) and divided into the two main types of hydrogen-poor Type I SNe and hydrogen-rich Type II SNe, in which all but sub-type Ia (thermonuclear explosion, see e.g. \citealt{maoz2014} for a review) are core-collapse SNe.

The hydrogen-poor Type I core-collapse SNe are further divided into sub-types Ib and Ic by features in the early spectrum around peak luminosity: Type Ib spectra show strong helium but no hydrogen absorption lines, Type Ic spectra have neither hydrogen nor helium lines. 

The hydrogen-rich Type II SNe are divided into four sub-types IIL, IIP, IIn and IIb. Spectra of the most abundant types IIP and IIL are characterised by broad hydrogen emission lines (e.g. \citealt{gutierrez2017}) indicating high expansion velocities of the SN ejecta. They are differentiated by the shape of their light curves: Light curves of Type IIP have a distinct plateau phase after the peak luminosity with almost constant or slowly declining luminosity, in contrast to the light curve of Type IIL showing a steady (more or less linear) decrease after the peak luminosity. Recent evidence does not support this strict separation between Type IIP and Type IIL but a continuous range of decay rates of Type II SNe (\citealt{anderson2014, sanders2015, galbany2016a, valenti2016}). Type IIn SNe, first introduced by \cite{schlegel1990}, are characterised by relatively narrow Balmer emission lines indicating strong interaction of SNe shock waves with dense circumstellar material (e.g. \citealt{taddia2013a}).  Type IIb SNe have an intermediate character between Type II and Type Ib SNe and exhibit hydrogen lines in the early, photospheric phase which disappear in the later nebular phase when the ejecta become optically thin (e.g. \citealt{fang2018}).

In addition to the regular core-collapse SN types, there are high energetic SNe events (“Hypernovae”) not fitting into the scheme above. They are generally classified as Type Ibc-pec, also named as Types Ic-bl (very broad lines in spectrum, e.g. \citealt{taddia2019}) or Type Ic-GRB (associated by a gamma ray burst; e.g. \citealt{woosley2006}). A class of its own are superluminous SNe with absolute magnitudes M\textsubscript{V} < -21 mag (e.g. \citealt{galyam2019, chen2021}).

The diversity of SNe types reflects the diversity of unknown parameters of SNe progenitors, with initial mass, age, metallicity and binarity as the most important parameters. Additionally, the nature of the progenitor has to explain the observed spectral lines of Type Ibc\footnote{throughout this paper “Type Ibc” means “Type Ib and Ic”, while “Ib/c” means a Type I core-collapse SN with ambiguous classification} SNe. An absence of hydrogen (and helium) lines suggest the loss of the outer envelopes of the progenitor star before the explosion, which is possible by two fundamentally different channels: either the progenitor is an evolved single massive (M$_{ZAMS}$ $\geq$ 25-30 M$_{\sun}$) Wolf-Rayet (WR) star (see \citealt{crowther2007} for a review) losing its outer envelopes by strong stellar winds (e.g. \citealt{kudritzki2000, vink2005, smith2014}), or it is a less massive star in a close binary system losing its outer envelopes to the companion by accretion (e.g. \citealt{podsiadlowski1992, eldridge2008, yoon2010, dessart2011}).

The most direct way to constrain the progenitor of a core-collapse SNe is to use archival images of the explosion site to identify the progenitor star. This approach has been successful for Type II SNe (e.g. \citealt{smartt2015, vandyk2017} and references therein). A statistically significant number of progenitor stars of the abundant Type IIP SNe have been identified as red supergiants (RSGs) with initial masses between ~8 M$_{\sun}$ and ~17 M$_{\sun}$ (\citealt{smartt2009, vandyk2017}). Type IIL SNe are quite rare but the few direct progenitor detections indicate an initial progenitors mass $\leq$ 25 M$_{\sun}$ (e.g. SN1996al, \citealt{benetti2016}; SN2009hd, \citealt{eliasrosa2011}; 2009kr, \citealt{elisarosa2010}) and all appear to have a low-mass H-envelope. A few direct detections of Type IIn progenitors have been reported (e.g. SN2005gl, \citealt{galyam2009}; SN2009ip, \citealt{foley2011}; SN2010jl, \citealt{smith2011a}), pointing to luminous blue variable stars as progenitors, but these detections are still ambiguous. The same ambiguity applies to Type IIb progenitor detections (e.g. SN1993J, \citealt{aldering1994, vandyk2002, maund2004}; SN2008ax, \citealt{crockett2008}; SN2011dh, \citealt{maund2011, vandyk2011}; SN2013df, \citealt{vandyk2014}; SN2016gkg, \citealt{tartaglia2017}), which suggest a variety of progenitors including RSGs or yellow supergiants, with or without a companion.

The direct detection of progenitors of Type Ibc SNe has been challenging and is complex. Binarity can play a crucial role in defining the SN subtype, however with pre-explosion imaging it is difficult to differentiate a low-mass binary system from a single high mass star. Consequently, the binary/single-star debate is unsolved for all direct detections of Type Ibc progenitors. One of the most studied and debated direct progenitor detections is the case of the Type Ib SN iPTF3bvn (\citealt{cao2013, groh2013, bersten2014, fremling2014, eldridge2015, kuncarayakti2015, eldridge2016, folatelli2016, hirai2017}) resulting in the still open question of whether the progenitor was a massive star or a close binary system. Most recently, \cite{kilpatrick2021} reported a pre-explosion image detection of the progenitor of the Type Ib SN2019yvr, inferring a cool and inflated progenitor, but again were unable to rule out a close binary scenario. The only direct detection of a Type Ic SN progenitor is for SN2017ein (\citealt{vandyk2018, kilpatrick2018}), however the results of these two studies are unable to distinguish between a high initial mass single star (M\textsubscript{ZAMS} ~ 47-55 M$_{\sun}$), a close binary with two high mass components (80+48 M$_{\sun}$), or a young compact star cluster. The study by \cite{xiang2019} additionally took advantage of very early (1-2 days after explosion) photometric and spectral data of SN2017ein and derived consistent constraints for the progenitor. In contradiction, \cite{teffs2021} derived a lower mass progenitor (16-20 M$_{\sun}$) from modelling of photospheric and nebular phase spectral data. Further observations of the SN2017ein explosion site are required to solve this disagreement and further constrain the progenitor. 

Many studies searching for Type Ibc progenitors failed with the direct detection but provided upper limits for the luminosity and mass of the progenitors (e.g. SN1994I, \citealt{barth1996, vandyk2016}; SN2000ew, \citealt{maund2005a}; SN2001B, \citealt{maund2005a}; SN2002ap,  \citealt{mazzali2002, smartt2002, crockett2007}; SN2004gt, \citealt{maund2005b, galyam2005}; SN2009jf, \citealt{valenti2011}; SN2012au, \citealt{pandey2021}; SN2012fh, \citealt{johnson2017}; SN2013dk, \citealt{eliasrosa2013}). Because the majority of these studies found upper luminosity limits too low for single massive WR stars, the binary formation channel is currently favoured for the stripped-envelope Type Ibc SNe. This is supported by arguments related to the stellar initial mass function (IMF) and the observed ratio of Type Ibc to Type II SNe (e.g. \citealt{smartt2009}). However, the binary channel for Type Ibc SNe is not definitely established and the single massive WR star cannot be ruled out (e.g. \citealt{smith2011b}).

Given the challenges of direct progenitor detection, other studies have investigated the variation of Type Ibc to Type II ratio with global properties of the host galaxies to constrain SN progenitors (e.g. \citealt{prantzos2003, boissier2009, arcavi2010, prieto2008, hakobyan2014, hakobyan2016}). However, the multiple stellar populations of the hosts with their different ages, metallicities, and star formation histories make it difficult to produce useful constraints on progenitors. To avoid such complications, most studies have tried to constrain progenitor properties from the local SN environment using either global proxies to derive the local parameters or by direct observations of the local SN environment (see \citealt{anderson2015} for a review). The motivation behind this approach is that a massive progenitor star can only travel a short distance from the place of birth to the observed explosion site due to its short lifetime. Therefore, age and metallicity of the stellar population and HII region at the explosion site should be representative of the SN progenitor itself.

Early work on SN environments by \cite{vandyk1992} found that approximately 50\% of SNe were associated with a HII region with no significant difference between Type II and Type I SNe. In contrast, \cite{anderson2008} found only a small fraction of Type II SNe associated with HII regions, whereas an association for Type Ibc was found. They used the pixel statistics technique (\citealt{fruchter2006, james2006}) with narrow-band H$\alpha$ images and concluded that Type Ibc progenitors are more massive than Type II (\citealt{anderson2008}). \cite{crowther2013} argued that there should be no association between Type II SNe and HII regions because the lifetime of typical HII regions are considerably shorter than the lifetime of RSGs. \cite{maund2018} evaluated the very young resolved stellar populations in the environments of stripped-envelope SNe and found decreasing characteristic ages of the populations from Type IIb, Type Ib to Type Ic (log(age) = 7.20, 7.05, and 6.57, respectively). The finding indicates a significant fraction of massive stars as progenitors of Type Ibc SNe. 

The metallicity of a SN progenitor is of particular interest because it determines – together with the luminosity – the strength of the stellar winds and consequently the mass loss of the progenitor (predicted mass loss rate $\dot{M} \propto$ Z\textsuperscript{0.42–0.85}, \citealt{vink2021}). Studies based on global proxies have used the radial position of the explosion sites together with the host central metallicities and the metallicity gradient of galaxies (e.g. \citealt{henry1999} for a review) to constrain the progenitor metallicity (e.g. \citealt{prieto2008, anderson2009, taddia2016}). However, studies with indirect measurement of metallicity at explosion sites derived from global proxies suffer from large uncertainties caused by e.g. galaxy interactions, recent galaxy mergers or unknown star formation history. Consequently, the direct measurement of gas metallicity at the local environments of SNe explosion sites is preferred.

Several studies (\citealt{modjaz2008, anderson2010, leloudas2011, modjaz2011, kelly2012, kuncarayakti2013b, kuncarayakti2013a, kuncarayakti2018, taddia2013b, taddia2015, galbany2016b, galbany2016c, galbany2018, kruhler2017, sanders2012, schady2019, xiao2019}) have measured the gas-phase metallicities of the HII regions at the SN explosion sites based on long-slit or integral field spectroscopy applying empirical strong emission line methods. Results of these studies are inconclusive and do not unambiguously constrain progenitor metallicities for  different types of SNe. The majority of studies concluded that environments of Type Ibc SNe tend to have higher metallicities than environments of Type II SNe, but the differences were not statistically significant.

Despite all the effort to constrain the physical parameters of SNe progenitors, most questions remain unanswered. This work presents metallicity data of the environments of a well-defined, volume-limited, large sample of SNe within a distance of 30 Mpc based on long-slit spectroscopic data and inferred by the strong emission line method. The paper is structured as follows: Section \ref{sect_obs} presents the SN sample selection criteria and the target observations. Section \ref{sect_dat} describes the data reduction process and the method of SN environment metallicity measurement. Section \ref{sect_res} presents the results followed by the discussion in Section \ref{sect_dis}. Finally, Section \ref{sect_con} summarizes the conclusions.

\section{Observations of SN sites} \label{sect_obs}

This project aims to evaluate differences in the metallicity of the environments of different core-collapse SN types based on a large sample of volume-limited SN detections. The sample was compiled according to selection criteria as follows: a) Luminosity distance $\lesssim$ 30 Mpc in order to get an appropriate spatial resolution and to avoid greater uncertainties of true local metallicity at higher redshifts, b) reliable spectral SN classification as Type IIP, Type Ib or Type Ic to avoid any skew by uncertain classifications and c) the host galaxy should have sufficiently low inclination (<75°) to avoid ambiguities of the SN HII region identification and to avoid issues with emission lines detection by high host extinction.

SNe with luminosity distances up to 33 Mpc were taken from the Open Supernova Catalogue\footnote{https://sne.space} (\citealt{guillochon2017}). All SNe up to 27 Mpc were accepted as target candidate. The luminosity distances of SNe between 27 Mpc and 33 Mpc were searched for additional distance measurements in the NASA/IPAC Extragalactic Database (NED\footnote{http://ned.ipac.caltech.edu}, \citealt{helou1991}) and accepted as candidate if distances <30 Mpc are consistently reported.

Classifications of the SNe were taken from the Open Supernova Catalogue. All SNe with clear Type Ib, Type Ic and Type IIP were accepted as targets. There is a large diversity in Type II subtypes (IIP. IIL, IIn, IIb) and \citet{Graham2019} presents work suggesting that different Type II subtypes have different metallicity distributions, although low number statistics for the rarer subtypes makes this difficult to confirm. To ensure this diversity does not affect our results we include only Type IIP SNe, which are abundant in number and have many confirmed RSG progenitors (\citealt{smartt2009}).

SNe with ambiguous classification as ‘Type Ib/c’ or just ‘Type II’ were examined by the references given in the Open Supernova Catalogue and assigned the subtype most commonly reported in the literature. Additionally, SNe older than 1990 were excluded because of uncertainties of the instrumentation and of the classification method used. SNe younger than 2017 were excluded to avoid a potential contamination of the HII region by (residual) light of the SN event itself.

Inclinations of the host galaxies were taken from Hyperleda\footnote{http://leda.univ-lyon1.fr}  (\citealt{makarov2014}). The inclination of the host galaxy is the least constraining criterion. The light of a SN environment in the outskirts of a high inclination galaxy may be less affected by stellar contamination than the light of a SN environment close to the centre of a low inclination galaxy. Thus, all SNe in hosts with inclination $<$75$^{\circ}$ were accepted as candidate and SNe in hosts with an inclination $>$75$^{\circ}$ were decided case by case depending on the location of the SN in the galaxy. This approach identified 40 Type Ib, 39 Type Ic and 107 Type IIP SNe as target candidates for the project.

\begin{table} %
	\caption{Overview of the five INT observations runs contributing observation data to the project}
    \begin{tabular}{p{0.5cm}p{2.4cm}ccp{0.8cm}}
    	\hline
    	\hline
    	year  & \multicolumn{1}{c}{nights} & \#nights & \#obs. & \multicolumn{1}{c}{INT prop. ID} \\
		\hline
		2016  & 27-Sep to 01-Oct & 5     & 25    & I/2016B/05 \\
		2017\textsuperscript{1}  & 27-Dec, 28-Dec & 2     & 14    & I/2017B/01 \\
		2018  & 18-Feb,\,13-Jun, 30-Oct,\,26-Nov & 4     & 8     & SI2018a02 \\
		2019  & 21-Feb, 22-Feb & 2     & 10    & SI2018a02 \\
		2019  & 23-Feb to 28-Feb & 6     & 39    & I/2019A/01 \\
    	\hline
    \end{tabular}%
	\begin{tablenotes}[para,flushleft] \footnotesize
		\item \textsuperscript{1}  Eight nights were allocated but the run was hindered by technical fault with the telescope for the first 6 nights for which additional discretionary director time was awarded under proposal ID S12018a02.
	\end{tablenotes}
  	\label{tab_int_runs}
\end{table}%

\begin{table}
\centering
    \begin{threeparttable}
	\caption{Technical configuration of IDS/EEV10 equipment for taking the calibration and science frames.}
    \begin{tabular}{ll}
		\hline
    	spectrograph & IDS \\
    	slit width & 1.5\arcsec \\
    	slit length & 3.3\arcmin \\
    	grating & R400V \\
    	central wavelength & 5802.4 {\AA} \\
    	dispersion & 1.41 {\AA}/px \\
    	resolving power ($\lambda/\Delta\lambda$)  & 1596@4500 {\AA} \\
    	resolution ($\Delta\lambda$)  & 2.87 {\AA} \\
    	filters & clear \\
    	detector & EEV10 \\
    	readout speed & slow \\
    	linearity & $\pm$0.2\% to 65535 ADU \\
    	saturation level & 65535 ADU \\
    	dark current & 4 e/hour @ 153 K \\
    	gain factor & 1.2 electrons/ADU \\
    	scale factor & 0.4\arcsec/px \\
    	\hline
    \end{tabular}%
    \label{tab_configuration}
    \end{threeparttable}
\end{table}%

Observational data were taken over 5 observational periods between September 2016 and February 2019 (Table \ref{tab_int_runs}) at the Isaac Newton Telescope (INT) on La Palma with the IDS/EEV10 instrument with the technical configuration shown in Table \ref{tab_configuration}. A total of 96 observations, of 76 individual SNe (Table \ref{tab_observations}), were obtained over 19 nights. Eleven targets were observed twice and two targets were observed three times to get an improved signal-to-noise ratio (SNR) and to check our methods for consistency/reproducibility. 

An additional four targets (SN1990aa, SN1991ar, SN1996D and SN2009ga), which are not part of the selected target sample, could be observed by opportunity. The data of these four targets were reduced and evaluated but their data are not incorporated in the statistical evaluation in Section \ref{sect_res} because the luminosity distance is significantly larger than 30 Mpc (lower uncertainty range limit >40 Mpc).

The observations include the standard calibration frames (bias and flat frames, arc frames for wavelength calibration, standard star frames for flux calibration). In order to remove cosmic rays, it was aimed to take three exposures for every target observation, which worked for all but three targets. Exposure time was typically 1200s per frame increased for faint targets and/or high sky brightness conditions up to 1800s per frame. Seeing conditions were between 0.8\arcsec to 1.2\arcsec for all observation nights.

\section{Data Reduction} \label{sect_dat}

The data reduction made use of the two standard astronomical software packages IRAF (Image Reduction and Analysis Facility; \citealt{tody1986}) and Starlink (\citealt{currie2014}). 

Standard data reduction processes were carried out to reduce and combine the 2-dimensional (2D) spectra. Due to a slight variation of the bias level during some nights, bias subtraction was done by means of the individual bias strip of the frames. Dark current correction was negligible because of the detector cooling. The data reduction included S-distortion correction and flux calibration was achieved using standard star observations taken throughout the night. The spectra were trimmed to 3500-7000 {\AA} given the location of the diagnostic lines required. 

The pixel position of the SN site was identified and extracted from the reduced 2D spectrum. Evaluation of the correct SN site requires evaluation of the slit-orientation, which may differ from the position angle (PA) on the sky by $\pm$180\textdegree. This was done by means of the INT acquisition images. With knowledge of the slit-orientation, the pixel row of the SN site spectrum was identified by the known angular distance of the SN site (Table \ref{tab_observations}) from host galaxy centre and the INT IDS spatial plate scale of 0.4\arcsec/px (Table \ref{tab_configuration}). Extraction of the SN site spectrum was done by the IRAF APALL task with an extraction aperture of 4\arcsec as a reasonable choice to account for seeing conditions, INT guiding accuracy and other imperfections in the optical path. The chosen 4\arcsec extraction aperture corresponds to a linear size of 19.4pc/Mpc or about 0.6 kpc for a host galaxy with luminosity distance 30 Mpc. 

\begin{figure*}
\includegraphics[trim={0cm 0cm 0cm 0cm}, clip, width=1.92\columnwidth, angle=0]{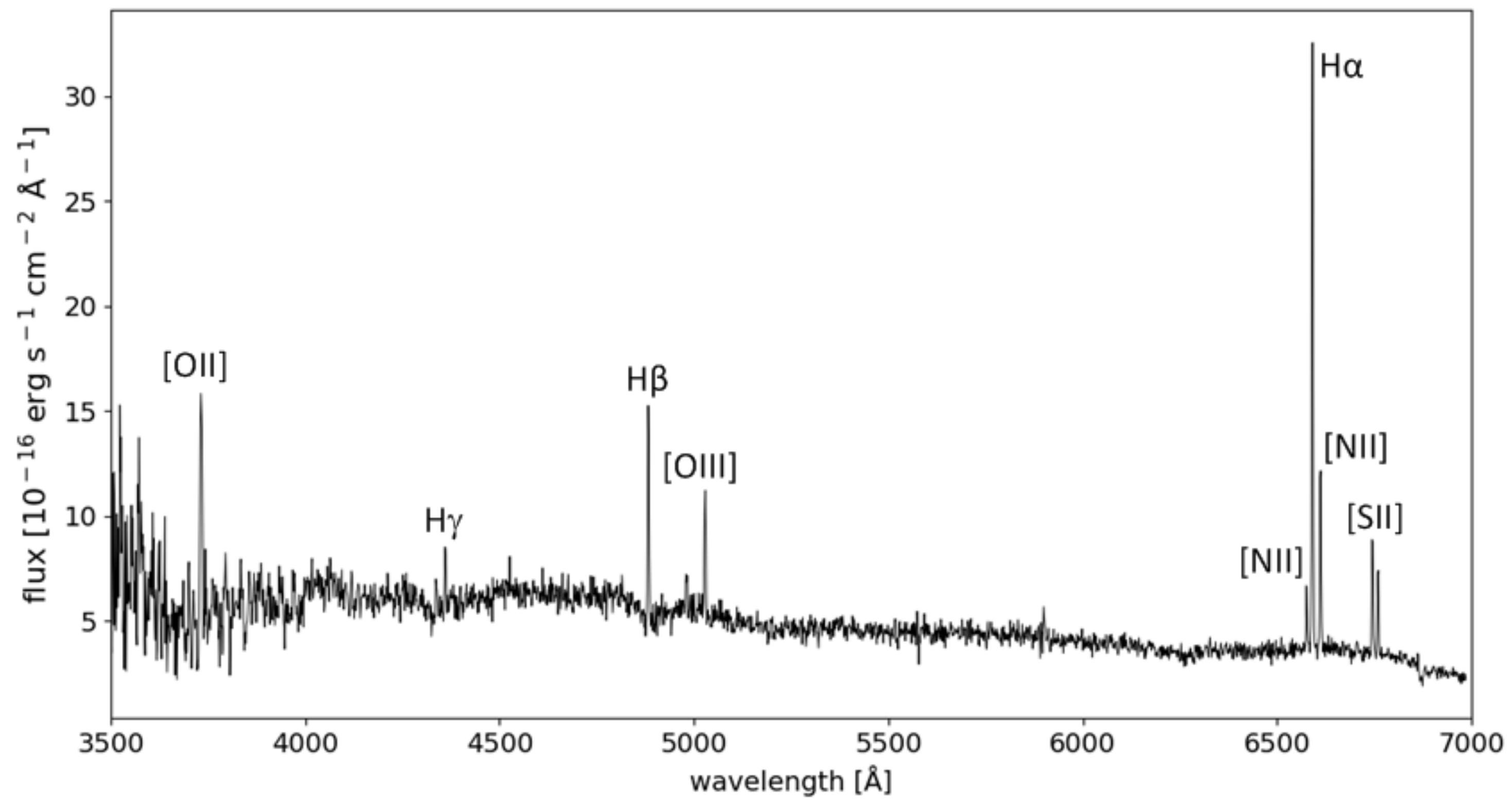}
\caption{Obtained INT spectrum of Type Ib SN2012au environment as an example for the extracted 1D environment spectra. The figure shows the most prominent emission lines of which H$\alpha$, [N\,{\sc ii}]$\lambda$6583, H$\beta$ and [O\,{\sc iii}]$\lambda$5007 have been used for the metallicity estimation.}
\label{fig_sn2012au}
\end{figure*}

Interstellar extinction for each SN region was derived from the Balmer line emission  based on a H$\alpha$/H$\beta$ ratio of 2.86, assuming case B recombination \citep{Hummer1987} and empirical extinction curves from \citet{osterbrock2006}. Line emission was measured using the Emission Line Fitting (ELF) routine in the Starlink package DIPSO and extinction correction was performed using the DRED routine. Target spectra lacking H$\beta$ emission were only corrected for Galactic extinction towards the SN host galaxy with values taken from NED (based on \citealt{schlafly2011}) and assuming a \citet{Fitzpatrick1985} reddening law. Figure \ref{fig_sn2012au} shows an example of the final extracted spectrum with the most prominent emission lines indicated.

\begin{figure}
\includegraphics[trim={0cm 0cm 0cm 0cm}, clip, width=1.0\columnwidth, angle=0]{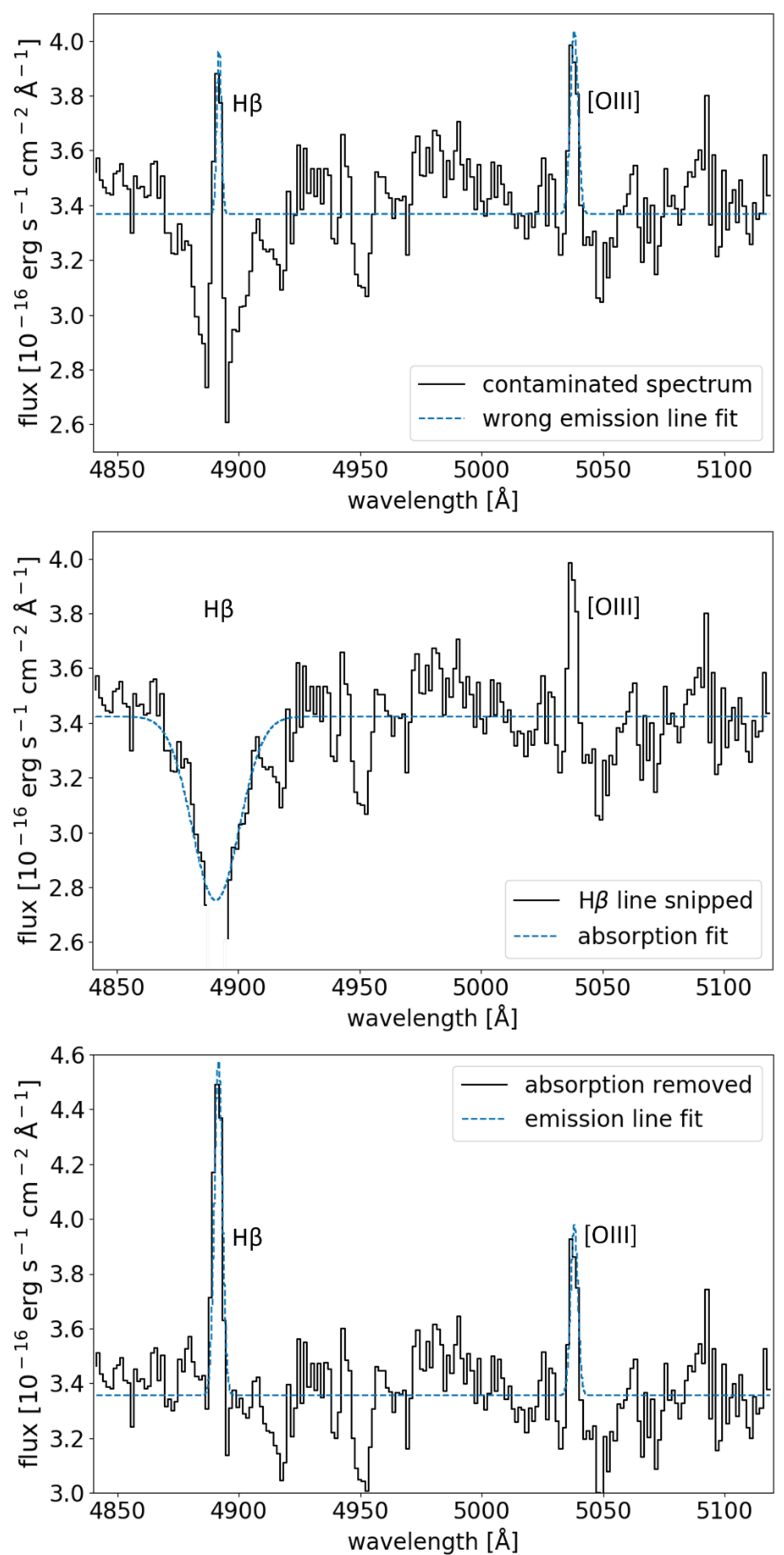}
\caption{Underlying stellar contamination may affect the H$\beta$ emission lines. This figure demonstrates the DIPSO workaround for the example SN2004fc as described in the text. The top plot shows the observed spectrum (black) of SN2004fc; the blue dashed line is the DIPSO ELF fit strongly affected by the contamination. The middle plot is the DIPSO fit to the absorption after snipping the H$\beta$ line out of the observed spectrum. The bottom plot shows the improved emission line fit after removal of the stellar contamination by applying the absorption fit from the middle plot.}
\label{fig_workaround}
\end{figure}

For flux calculation, the emission lines were fitted by Gaussian profiles using the DIPSO command ELF. The ELF fits have been done in the H$\alpha$ and in the H$\beta$ region independently, but the fits within each region, i.e. H$\alpha$/[NII] lines and H$\beta$/[OIII] lines, respectively, have been done simultaneously. From all observations, 10 targets were affected by strong stellar contamination clearly visible by underlying broad absorption features at the H$\beta$ emission line. The ELF command is not able to disentangle the emission lines flux from this stellar absorption at the same wavelength. To account for this absorption the fitting has been modified as follows: a) cut the emission line affected by stellar contamination from the spectrum; b) fit the absorption feature by ELF and c) subtract the Gaussian fit of the absorption feature from the initial spectrum, leaving just the emission line (see example in Figure \ref{fig_workaround} for target SN2004fc). This worked well for all affected targets with strong stellar contamination and as long as the underlying absorption line is broader than the emission line. If the stellar absorption has a width less than or equal to the emission line width, then the workaround is not applicable because the emission line masks the absorption (see discussion in Section \ref{sub_con}). A stellar contamination of the H$\alpha$ lines was not visible at all and estimated as negligible compared to the strength of H$\alpha$ line (see also Section \ref{sub_con}).

The project made use of the empirical strong emission line method for measuring the metallicity of SNe environments, because the auroral [O\,{\sc iii}]$\lambda$4363 line required for the direct method is weak and often undetected. The strong emission line method uses only the flux ratios of strong nebular emission lines, which are much easier to detect, for the metallicity determination. The methods provide the ratio of the number density oxygen to hydrogen (in terms of 12$+$log(O/H)) as a proxy for metallicity.

Many different ratios of strong emission lines have been used to determine gas-phase metallicities (see e.g. \citealt{kewley2008, kewley2019} and references therein). The empirical approach of the method requires a calibration with a large number of H\,{\sc ii} regions, for which both the inferred metallicities by strong emission lines and the metallicity derived by the direct method, are known. This work uses the N2 and O3N2 methods introduced by \cite{pettini2004} with the updated calibration by \cite{marino2013}, where the N2 indicator is defined by
\begin{equation}
\text{N2} = \log \left( \frac{\text{[NII]}\lambda\text{6583}}{\text{H}\alpha}\right) 
\label{N2_indicator}
\end{equation}
and the O3N2 indicator is defined by
\begin{equation}
\text{O3N2} = \log \left( \frac{\text{[OIII]}\lambda\text{5007/H}\beta}{\text{[NII]}\lambda\text{6583/H}\alpha} \right). 
\label{O3N2_indicator}
\end{equation}
The N2 and O3N2 methods have been chosen because they use closely spaced emission line ratios making the methods robust to reddening and flux calibration issues. Additionally, these two methods are consistent with spectroscopically derived massive star abundances (\citealt{davies2017}).

\section{Results}   \label{sect_res}

The H$\alpha$ and [N\,{\sc ii}]$\lambda$6583 emission lines required for the N2 indicator were detected in 66 of the 76 observed SNe environments. The O3N2 method additionally needs H$\beta$ and [O\,{\sc iii}]$\lambda$5007 lines, which were detected in 46 observations. Metallicity estimation was not possible for 10 observed targets because either no emission lines were detected (SN1995bb, SN2003ie, SN2004ez, SN2005cz, SN2007od) or only hydrogen lines were present (SN1999ev, SN2005ad, SN2010br, SN2015aq, SN2017iro). We note that these SNe are a mix of SN subtypes. 

Table \ref{tab_results} shows the measured metallicities of the SN environments inferred by N2 and O3N2 (where applicable) methods. M13-N2 and M13-O3N2 are the results for the calibration by \cite{marino2013} accordingly calculated by the equations 
\begin{equation}
\text{M13-N2} = 8.743 + 0.462 \times \text{N2} 
\end{equation}
and
\begin{equation}
\text{M13-O3N2} = 8.533 - 0.214 \times \text{O3N2}
\end{equation}

For compatibility with previous studies, the columns PP04-N2 and PP04-O3N2 present the results with the initial calibration by \cite{pettini2004}, calculated by
\begin{equation}
\text{PP04-N2} = 8.90 + 0.57 \times \text{N2} 
\end{equation}
and
\begin{equation}
\text{PP04-O3N2} = 8.73 - 0.32 \times \text{O3N2}
\end{equation}

The statistical uncertainties resulting from the calibration itself are $\pm$0.16 dex for M13-N2, $\pm$0.18 dex for M13 O3N2 (1$\sigma$ values, \citealt{marino2013}), $\pm$0.18 dex for PP04-N2 and $\pm$0.14 dex for PP04-O3N2 (1$\sigma$ values, \citealt{pettini2004}). The observational uncertainties of our observed data by the photon noise and the data reduction process have been estimated to between $\pm$0.01 and $\pm$0.04 dex mainly depending on the SNR. This uncertainty is significantly less than the calibration uncertainties, and so has been neglected for simplicity. 

For completeness, Table \ref{tab_opportunities} shows the metallicities of the environments of SN1990aa, SN1991ar, SN1996D and SN2009ga observed by opportunity. These 4 `opportunity' targets do not show any remarkable difference to the metallicities of Table \ref{tab_results}.

The N2 metallicities for 14 Type Ib, 19 Type Ic and 33 Type IIP SN environments are obtained and O3N2 metallicities have been derived for 9 Type Ib, 18 Type Ic and 19 Type IIP SN environments, and are presented in Table \ref{tab_results}. SN2000ds, a Ib SN, has been excluded from the statistical evaluation as it has been classified as a Ca-rich SN (e.g. \citealt{filippenko2003}). Such events are predicted to be part of the Ca-rich gap transients which are a different class of transients often found on the outskirts of elliptical galaxies suggesting an old (and hence low mass) progenitor population \citep{lyman2013,lunnan2017,shen2019}. This leaves 13 Type Ib SN available for statistical evaluation with the N2 indicator; we note that no O3N2 result is available for SN2000ds.

Table \ref{tab_statistics} lists the mean and standard deviation of  metallicities for different SN types and for the total sample. Differences in the mean metallicities for each subtype are small, with the largest difference being between Ib and IIP SN at 0.08\,dex for the O3N2 calibration, but this is not statistically different if the standard error on the mean is considered and is not seen in the N2 calibration results. However, both calibrations indicate a large difference between the standard deviation of Type Ib SN compared to the two other subtypes. There is a factor of 2.0 (2.5) between Type Ib and both Type IIP/Ic standard deviations for the N2 (O3N2) calibrations, which is remarkable. The implications of this result are discussed in Section \ref{sub_con}.

\begin{table}
\caption{Number of targets (N), mean values and standard deviations ($\sigma$) of the metallicities split into the three SN types and for the total sample based on the M13-N2 and M13-O3N2 calibration results.}
    \begin{tabular}{c|ccc|ccc}
    \hline
    \hline
    SN    & \multicolumn{1}{c}{N(N2)} & \multicolumn{2}{c|}{M13-N2} & \multicolumn{1}{c}{N(O3N2)} & \multicolumn{2}{c}{M13-O3N2} \\
    type  &       & \multicolumn{2}{c|}{[12+log(O/H)]} &       & \multicolumn{2}{c}{[12+log(O/H)]} \\
    \multicolumn{1}{c|}{} &       & \multicolumn{1}{c} {mean} & \multicolumn{1}{c|}{$\sigma$} &       & \multicolumn{1}{c} {mean} & \multicolumn{1}{c}{$\sigma$} \\
    \hline
    Ib    & 13    & 8.52  & 0.07  & 9     & 8.50  & 0.06 \\
    Ic    & 19    & 8.49  & 0.14  & 18    & 8.49  & 0.16 \\
    IIP   & 33    & 8.52  & 0.14  & 19    & 8.42  & 0.15 \\
    \hline
    all   & 65    & 8.51  & 0.13  & 46    & 8.46  & 0.14 \\
    \hline
    \end{tabular}%
  \label{tab_statistics}%
\end{table}%

\begin{figure*}
\subfigure{\includegraphics[trim={0.3cm 0cm 1.35cm 1.1cm}, width=1.0
\columnwidth]{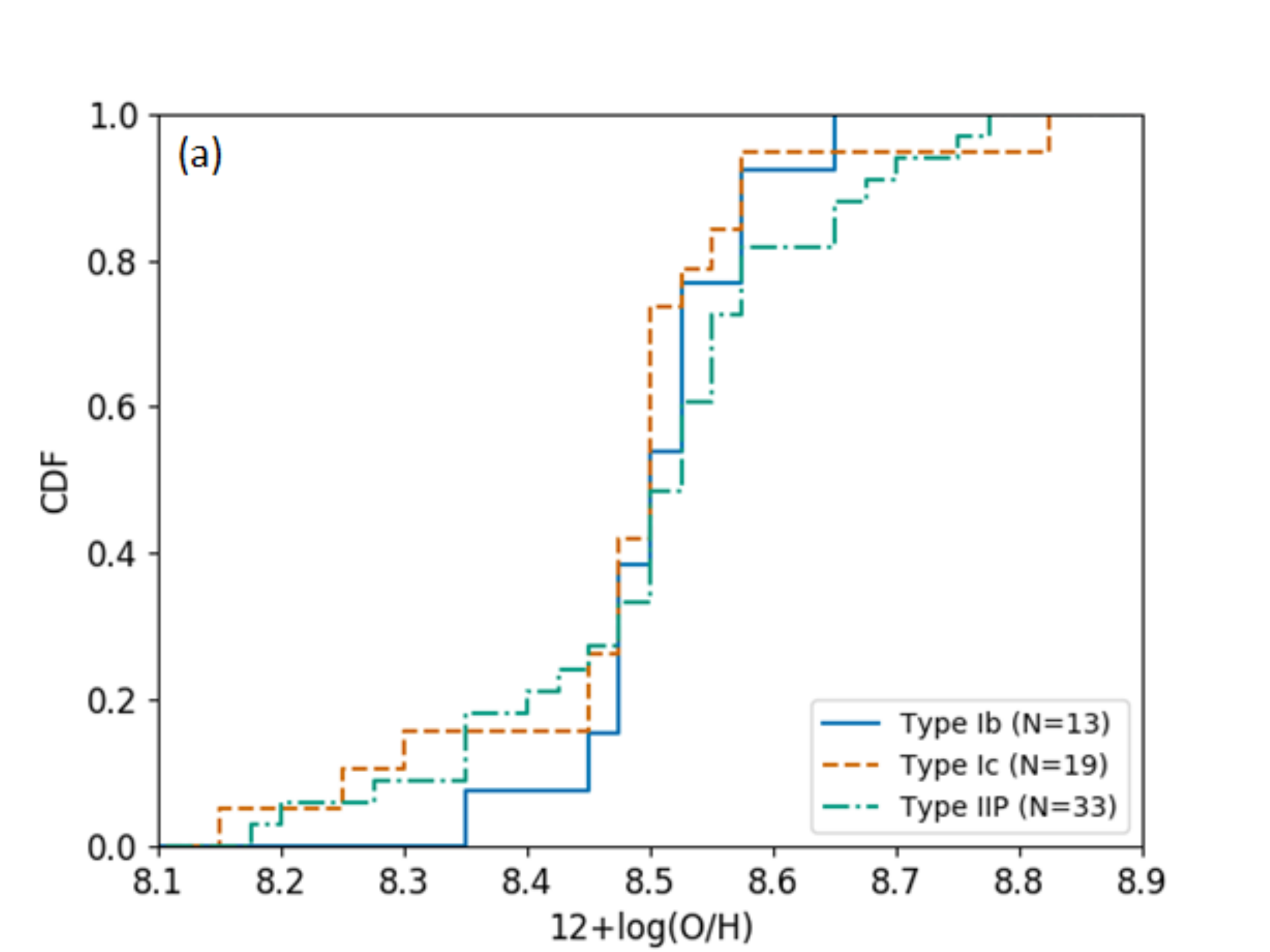} \label{fig_n2_cdf}}
\subfigure{\includegraphics[trim={0.3cm 0cm 1.35cm 1.1cm}, width=1.0\columnwidth]{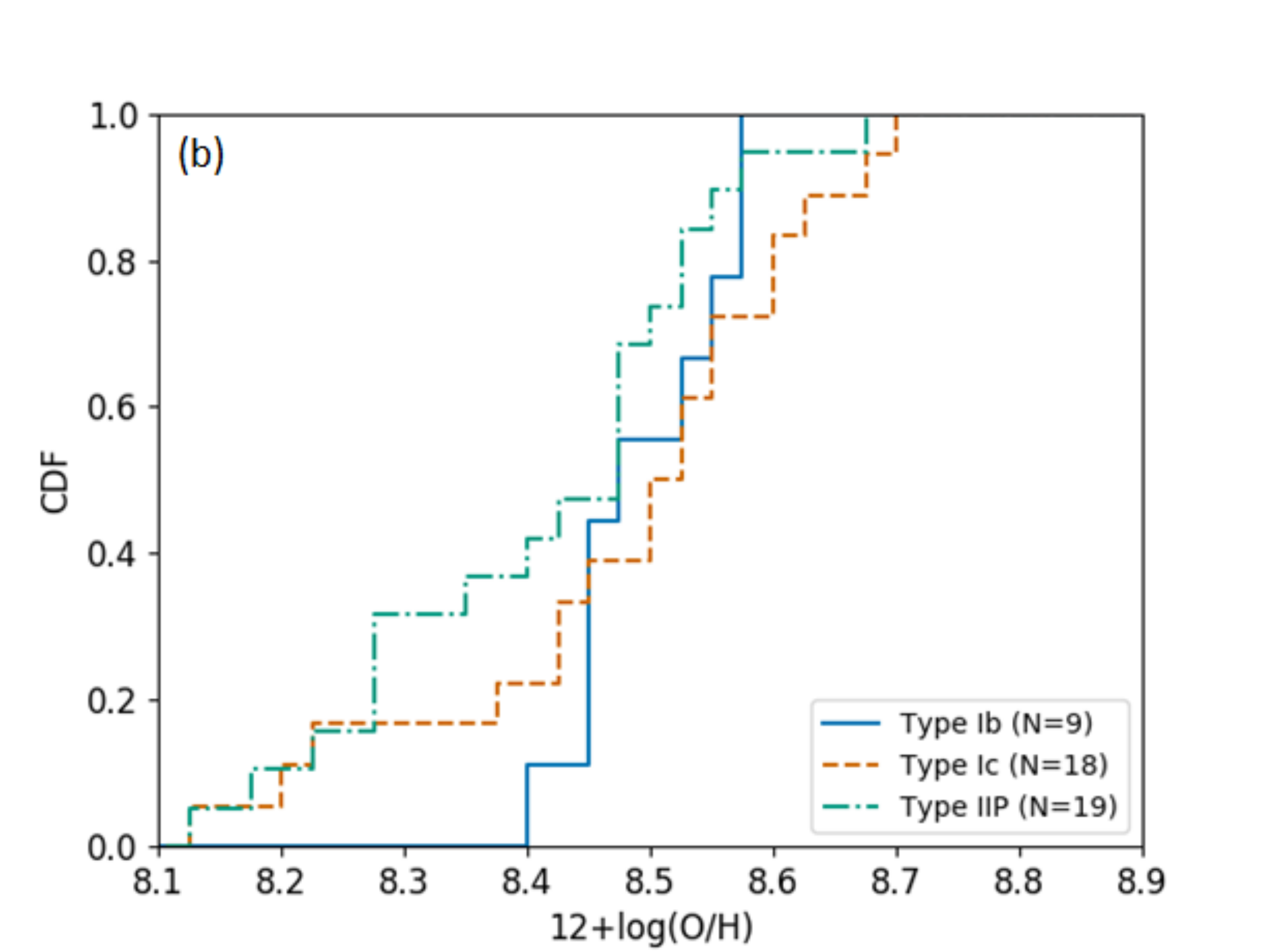} \label{fig_o3n2_cdf}}
\caption{CDFs of the SN environment metallicities measured with the M13-N2 (left) and M13-O3N2 (right) calibration. Binning width for CDF calculation: 0.025 dex.}
\label{fig_cdfs}
\end{figure*}

Figure \ref{fig_cdfs} shows the normalised cumulative distribution functions (CDFs) of the results for M13-N2 and M13-O3N2 calibrations. Reflecting the results presented in Table \ref{tab_statistics}, there seem to be two clear tendencies in the CDF estimated by M13-O3N2 calibration: firstly, the mean metallicity of Type Ibc environments show a tendency to be higher (0.08\,dex) than that of Type IIP and, secondly, the distribution of Type Ib is significantly narrower than the distribution of the two other types, indicating a restricted metallicity range for Ib SNe. (Type Ib standard deviation 0.06 dex compared to 0.16\,dex; see Table \ref{tab_statistics}). The first tendency above is not visible in the M13-N2 calibrations but the narrower distribution of Type Ib is still present.

In order to evaluate for statistically significant differences between the samples of the different SN-types, two statistical tests have been applied: the two-sample Kolmogorov-Smirnov test (KS-test) and the two-sample Anderson-Darling test (AD-test). Both tests are applicable for small samples (see e.g. tables in \citealt{massey1952} and \citealt{pettitt1976}).

\begin{figure*}
\includegraphics[trim={0cm 0cm 0cm 0cm}, clip, width=1.92\columnwidth, angle=0]{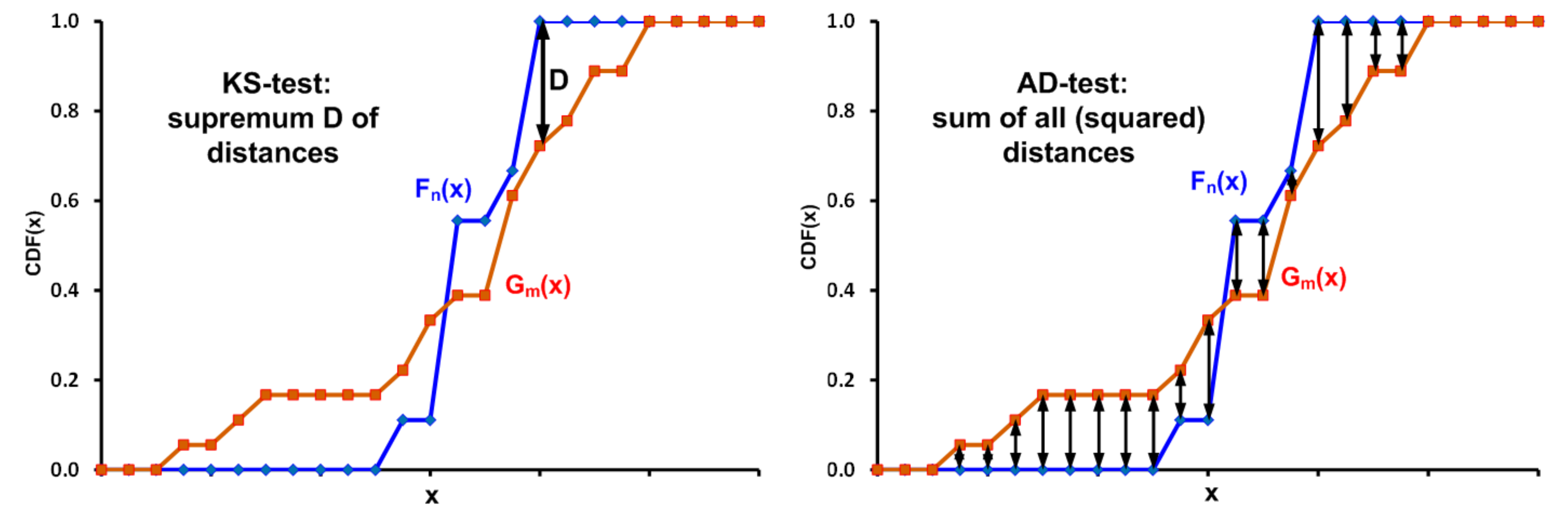}
\caption{Approach of the two-sample KS-test (left) and of the two-sample AD-test (right) to test the null hypothesis that two samples are drawn from same parent population. The KS-test uses only the supremum of absolute values of the distances between two CDFs to test the null hypothesis. The AD-test uses the sum of all (squared) distances between two CDFs to judge the null hypothesis. Compared with KS-test, the AD-test is more sensitive to CDF differences in the tails.}
\label{fig_ks_and_ad}
\end{figure*}

The tests have different approaches (schematically shown in Figure \ref{fig_ks_and_ad}) to test the null hypothesis that two discrete samples are drawn from same parent population. The two-sample KS-test (e.g. \citealt{press1988}) calculates differences between two discrete samples by evaluation of the supremum D = sup\textsubscript{x}$\vert$F\textsubscript{n}(x)- G\textsubscript{m}(x)$\vert$ of the distances between two CDFs F\textsubscript{n}(x) and G\textsubscript{m}(x) (Figure \ref{fig_ks_and_ad} left) of two samples with sizes n and m, respectively. If the value of the supremum exceeds a critical value D\textsubscript{crit}, the null hypothesis will be rejected and the two samples do not have the same parent population at certain significance. The value of D\textsubscript{crit} is calculated by the Kolmogorov distribution (\citealt{marsaglia2003}) and depends on the significance level $\alpha$ (usually 5\% or 1\%) and the sizes n and m of the two samples. 

While the KS-test evaluates just the supremum of the distances between the CDFs of two samples, the AD-test takes into account all distances between the two CDFs (Figure \ref{fig_ks_and_ad} right). Consequently, differences in the tails of the CDFs are more weighted and the AD-test judges more the area between the CDFs than just a maximum value. Mathematically, the two-sample AD-test calculates a value $\text{A}^2_\text{nm}$  by:
\begin{equation}
\text{A}^2_\text{nm} = \frac{\text{nm}}{\text{N}} \int_{-\infty}^{\infty} \frac{\left[\text{F}_{\text{n} }\text{(x)} - \text{G}_{\text{m}}\text{(x)}\right]^{2}} {\text{H}_{\text{N}}\text{(x)}\left[\text{1} - \text{H}_{\text{N}}\text{(x)}\right]} \text{ dH}_{\text{N}}\text{(x)}
\end{equation}
(\citealt{pettitt1976}) where F\textsubscript{n}(x) and G\textsubscript{m}(x) are the CDFs of the two samples with sizes n and m, N = n + m and H\textsubscript{N}(x) is the CDF of the combined sample of the two samples given by: H\textsubscript{N}(x) = [nF\textsubscript{n}(x) + mG\textsubscript{m}(x)]/N (\citealt{pettitt1976}).

The two-sample KS-test and AD-test are implemented in this work using the package `twosamples'  of the statistical software called `R'\footnote{https://www.r-project.org} \citep{Rproject2018}. The statistical tests of the package `twosamples' are not based on tables to calculate the p-value of the given samples but perform real permutations between samples to evaluate the p-value. The null hypothesis, that both samples have same parent population, must be rejected if the p-value is less than the chosen significance level $\alpha$.

\begin{table*}
  \begin{threeparttable}
	\caption{P-values of the two-sample KS-test and two-sample AD-test for the M13 results calculated by the functions `ks\_test' and `ad\_test’ of the R-project package `twosamples'. All p-values exceed the chosen significance level of 5\% and the null hypothesis that the two samples are drawn from same parent population cannot be rejected.}
    \setlength{\tabcolsep}{0.151cm}
    \begin{tabular}{c|cc|cc}
    	\hline
    	\hline
    	& \multicolumn{2}{c|}{KS-test} & \multicolumn{2}{c}{AD-test} \\
   		\hline
    	SN type & M13-N2 & M13-O3N2  & M13-N2 & M13-O3N2 \\
 		 & p-value & p-value  & p-value & p-value \\
    	\hline
    	Ib-Ic & 0.766 & 0.704  & 0.669 & 0.483 \\
   		Ib-IIP & 0.723  & 0.128 & 0.449 & 0.137 \\
  		Ic-IIP & 0.278 & 0.309 &  0.452 & 0.258 \\
   		\hline
        Ib-(Ic+IIP) & 0.749 & 0.311 & 0.583 & 0.279 \\
        \hline
  \end{tabular}%
  \label{tab_m13testresults}%
  \end{threeparttable}
\end{table*}%

Table \ref{tab_m13testresults} lists the p-value results of the sample combinations Ib-Ic, Ib-IIP and Ic-IIP for M13-N2 and M13-O3N2 metallicity calibrations for both the two-sample KS-test and the two-sample AD-test. The p-value for a significance level of 5\% must be <0.05 to reject the null hypothesis. The p-values in Table \ref{tab_m13testresults} are all significantly greater than 0.05 and thus the null hypothesis is not rejected. The KS-test and AD-test detect no statistically significant difference of the SN environment metallicity between the different SN subtypes. Table \ref{tab_m13testresults} also shows the p-values of the test of Type Ib with the combined sample Ic+IIP. If the parent populations of Type Ic and Type IIP are the same, then the p-value of the combined sample test should be lower than the p-values of the single samples Ic and IIP, respectively. This is not the case; the test provides no statistical evidence that Type Ic and Type IIP have same parent population in agreement with observational findings.  

\section{Discussion}  \label{sect_dis}

The reliability of the results presented in Section \ref{sect_res} has been investigated by several measures: impact of stellar contamination,  reproducibility of the results, reliability of the N2 and O3N2 calibrations, and effect of the metallicity uncertainties on the final p-value have been investigated using Monte-Carlo (MC) simulations of the KS- and AD-test.

\subsection{Stellar Contamination}    \label{sub_ste}

As discussed in Section \ref{sect_dat} a HII region spectrum may be contaminated by stellar radiation of stars within the HII region and/or stars along the line of sight. This stellar contamination especially affects the hydrogen lines resulting in an underlying absorption impacting the true emission line flux.

The DIPSO workaround applied to remove stellar contamination does not work if the stellar absorption has a width less than or equal to the emission line width. For this reason the metallicity results of Table \ref{tab_results} have been checked independently by the more sophisticated penalized PiXel Fitting (pPXF) method of emission line fitting (\citealt{cappellari2004, cappellari2017}). Using synthetic spectral templates of stars with varying parameters (available e.g. from the MILES stellar population library\footnote{http://research.iac.es/proyecto/miles/pages/ssp-models.php}, \citealt{vazdekis2010}) the observed continuum spectrum is fit allowing stellar kinematics and population parameters as well as the emission line flux to be extracted. The tests with pPXF reveal stellar contamination for most targets with H$\beta$ emission lines flux generally more strongly affected than H$\alpha$ emission (as expected for a contamination by a young massive star population). However, the derived metallicities based on the fluxes from pPXF line fitting and DIPSO line fitting are consistent within $\pm$0.03 dex in the majority of targets. For 7 out of 65 M13-N2 measurements the metallicities differ significantly ($>$0.1 dex), however this is mainly because of numerical issues of pPXF and/or DIPSO caused by low SNR observations. The differences between the pPXF and DIPSO metallicities of these 7 measurements are well within the uncertainty given by the calibration and do not affect the global M13-N2 results.

\begin{figure*}
\begin{minipage}[t]{0.99\columnwidth}
\includegraphics[trim={0.7cm 0cm 0.5cm 0cm}, clip, width=0.99\columnwidth]
{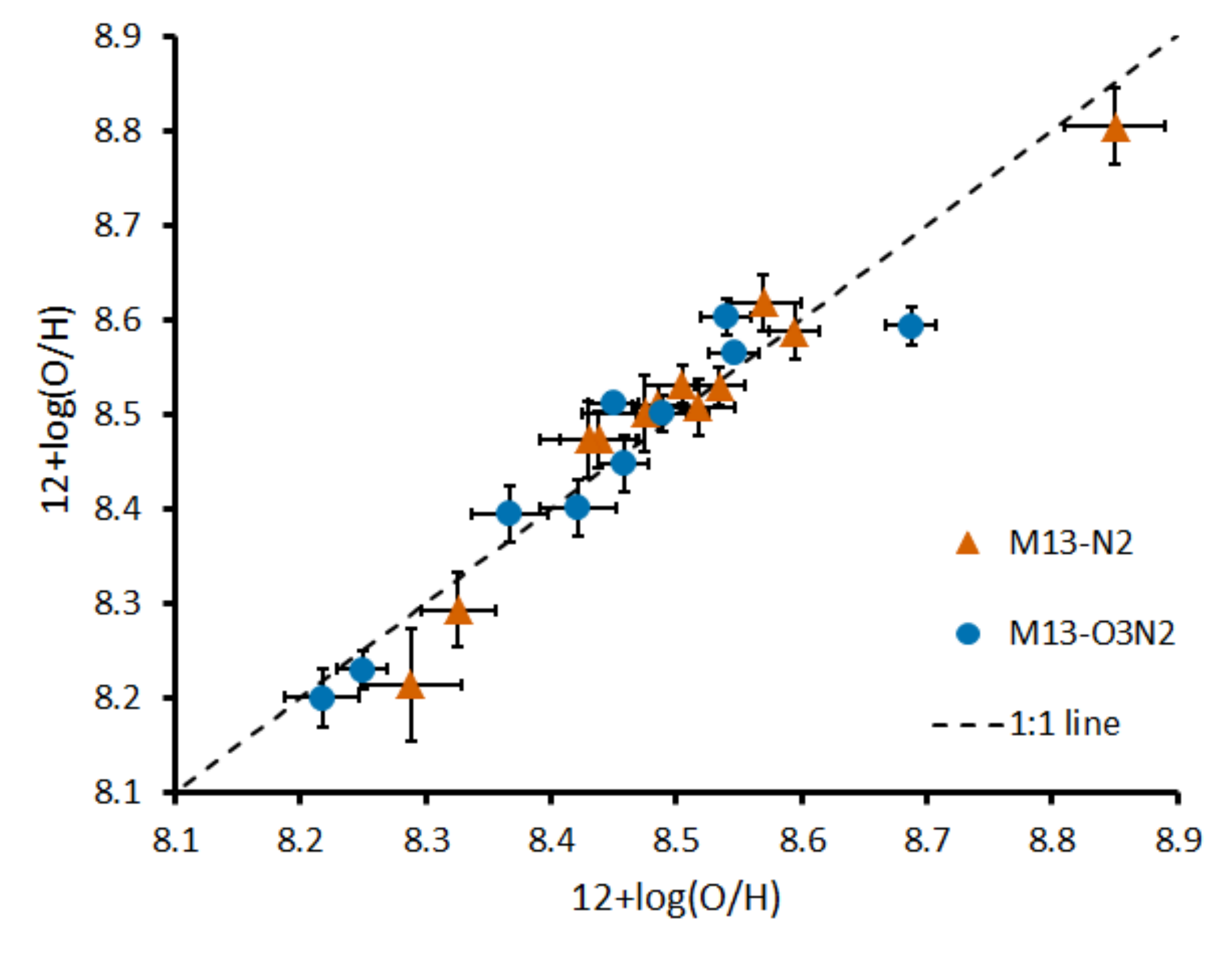}
\caption{Scatter plot of INT targets observed multiple times. The figure shows metallicities of 1\textsuperscript{st} observations vs. 2\textsuperscript{nd} observations. For targets with three observations, the worst case difference is shown. The time interval between the two observations was at least eight months. The error bars indicate the uncertainties caused by photon noise and data reduction process only and do not include the calibration uncertainties, which dominate the overall uncertainty (see Table \ref{tab_results}). The dashed line represents the hypothetical 1:1 line expected for identical values.} 
\label{fig_mult_obs}
\end{minipage}
\hspace{.09\columnwidth}
\begin{minipage}[t]{0.99\columnwidth}
\includegraphics[trim={0.3cm 0.0cm 1.8cm 1.5cm}, clip, width=1.0\columnwidth]{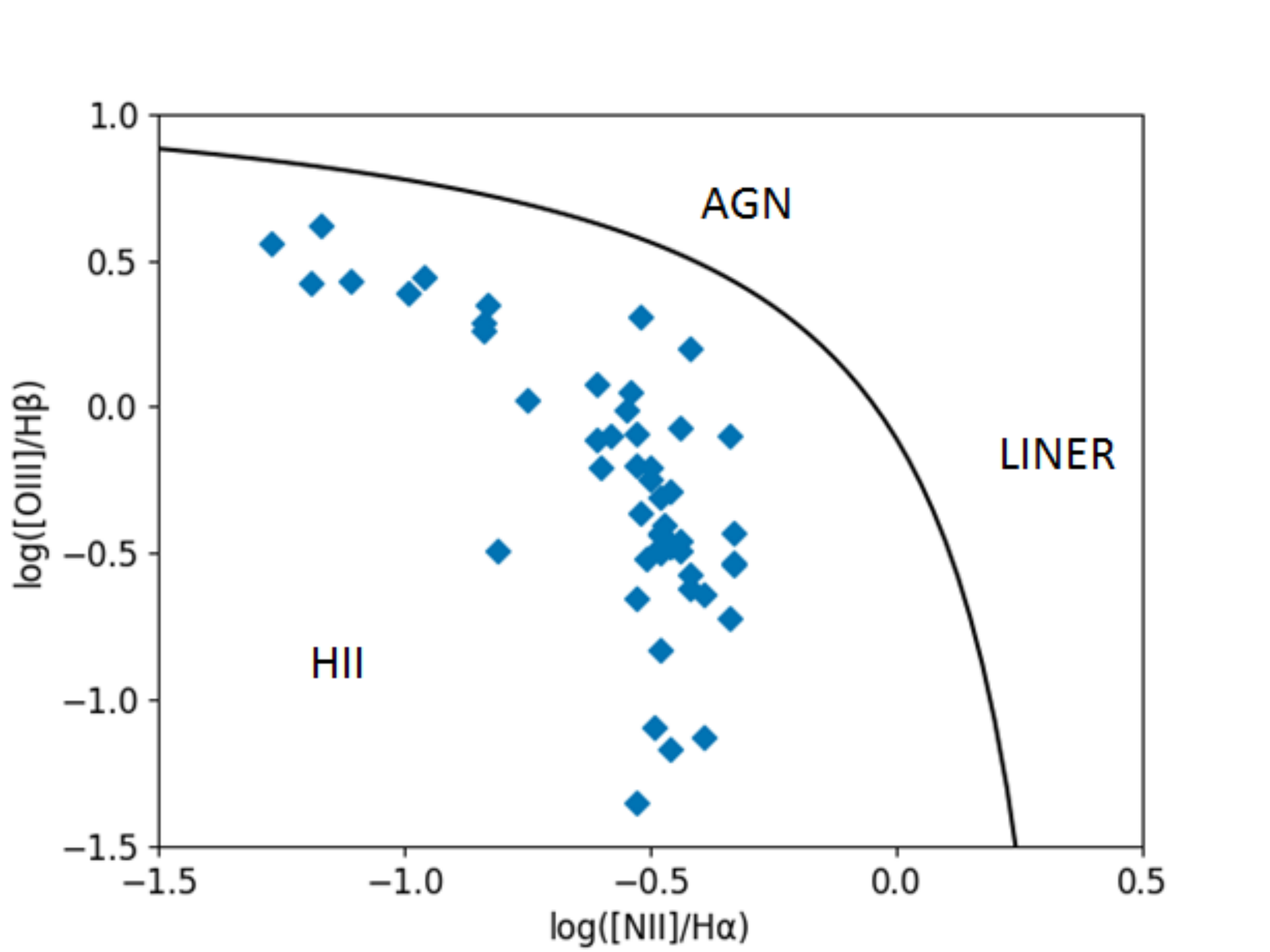}
\caption{BPT-diagram of all targets with O3N2 calibration (blue diamonds). The black solid line is the decision line between the HII region and the AGN/LINER region as given by \protect\cite{Kewley2001}, equation (5). All targets are well within the HII region of the BPT-diagram}
\label{fig_bpt_diagram}
\end{minipage}
\end{figure*}

\begin{figure}
\includegraphics[trim={0.4cm 0cm 0.5cm 0cm}, clip, width=0.99\columnwidth]{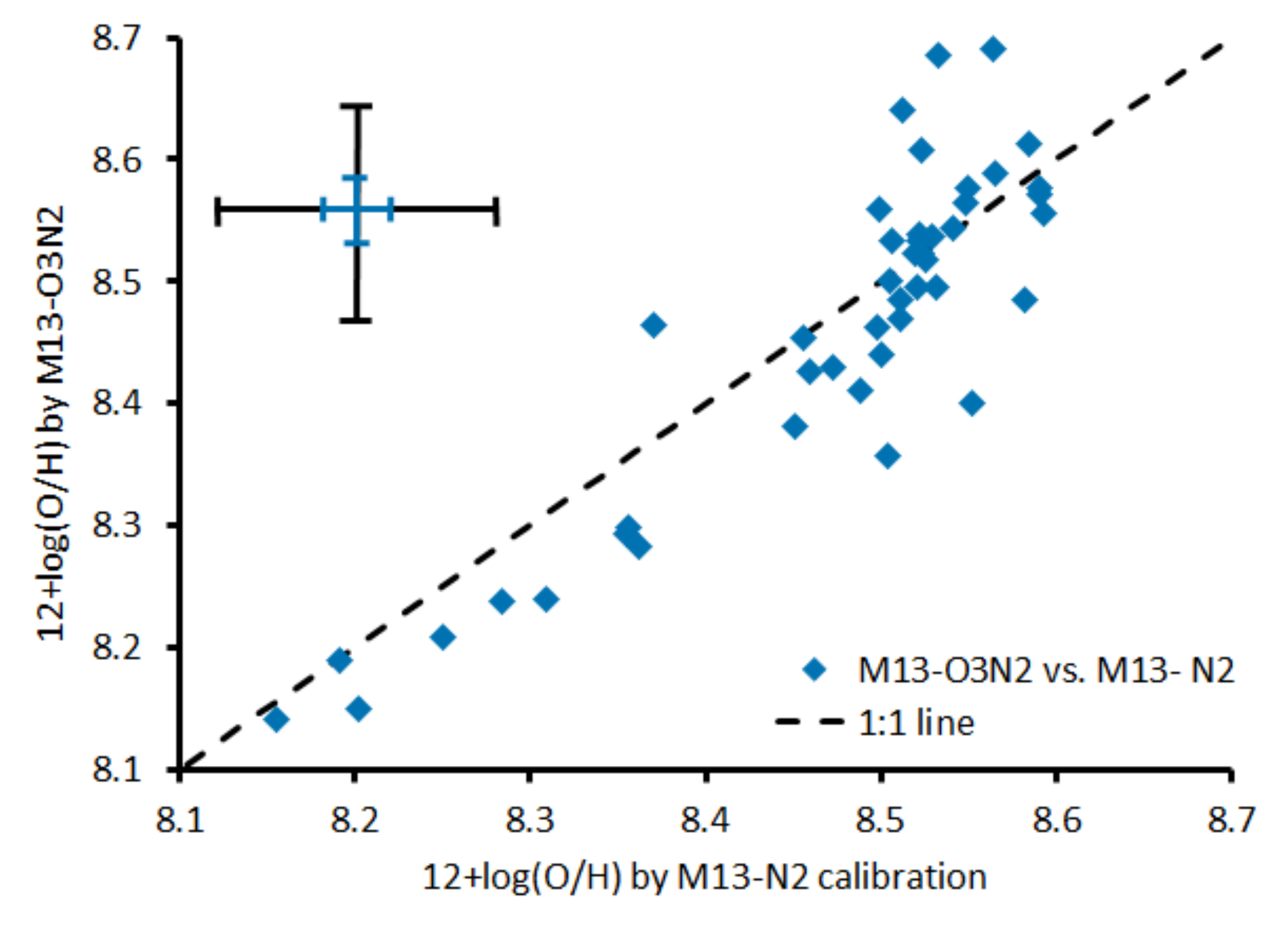}
\caption{Comparison of metallicity for SN where results using both the M13-O3N2 and M13-N2 calibrations is available. The black cross at top left indicates the overall uncertainty of the metallicities dominated by the calibration uncertainties; the overlaid blue cross indicates the typical observational uncertainty. The dashed line represents the hypothetical 1:1 line expected for identical values. There is an apparent pileup of points for x-axis values > 8.5\,dex, where the scatter significantly increases.}
\label{fig_scatterplot}
\end{figure}

\subsection{Reproducibility}    \label{sub_rep}

There are 11 targets observed twice and two targets observed three times during the project allowing us to evaluate the reproducibility of results. Figure \ref{fig_mult_obs} compares the metallicity for the multiple observed targets. The differences are all but one within the observational uncertainty of $\sim$0.04\,dex caused by the photon noise and the data reduction process. The larger difference of 0.08\,dex for SN2002jz are likely caused by very different seeing conditions (1.2\arcsec vs. 0.8\arcsec) between the two observations in 2016 and 2019. 

The consistency of results over different observing runs for the same SNe gives us confidence in the results of the relative metallicities and subtype distributions.

\subsection{Reliability of N2 and O3N2 results}    \label{sub_rel}

The application of the strong emission lines method is restricted to gas phase emission caused by stellar ionisation. We checked this precondition by means of the BPT-diagram \citep{baldwin1981} shown in Figure  \ref{fig_bpt_diagram}. All targets with O3N2 results are well within the HII region; for targets with N2 results only, the BPT diagram is not applicable.

The mean and standard deviations for the N2 and O3N2 calibrations, presented in Table \ref{tab_statistics} exhibit small quantitative differences between the two calibration methods. Firstly, the standard deviations of Type Ib differ from the other two SN subtypes by 0.07\,dex for N2 and 0.1\,dex for O3N2, and secondly the mean metallicities determined from the O3N2 calibration for the Types Ibc are both $\sim$0.08\,dex larger than the Type IIP mean metallicity, however no differences are identified using the N2 calibration. It is unclear why the two calibrations present different results. It could be caused by the larger number of N2 results (65 N2 results compared with 46 O3N2 results), by differences between the two calibration methods or an indication of real distinctions between the environments of different SN types that are only detectable with the O3N2 calibrator as a result of using more spectral information than the N2 calibration. Investigating the source of these differences is beyond the scope of this work and requires additional data.

The N2 and O3N2 calibrations (equations 3 to 6) are derived from empirical data and as such each calibration is only valid over a certain range. The validity ranges for the N2 and O3N2 indicators from \citet{marino2013} and \citet{pettini2004} are converted into terms of absolute metallicity and presented in Table \ref{tab_validity_limits}. Comparison of the SN metallicity results in Table \ref{tab_results} with the corresponding validity range of M13-O3N2 and PP04-O3N2 calibrations show that no SN lies outside of the validity interval when taking into account the calibration uncertainties of $\pm$0.18\,dex and $\pm$0.14\,dex, respectively

For the M13-N2 calibration eight SN lie outside of the validity range, however all but two (SN2000ds and SN2012cw) are within the metallicity calibration uncertainty of $\pm$0.16\,dex. Nine SN metallicities determined using the PP04-O3N2 calibration lie outside its validity range and four exceed the upper validity limit even when taking into account the $\pm$0.18\,dex calibration uncertainties. These SN are SN2000ds and SN2012cw, consistent with M13-N2, and also SN1999em and SN2014bc. SN2014bc is very close (about 160pc) to the active galactic nuclei (AGN) of its host NGC\,4258 and the spectrum may be contaminated by non-thermal radiation from the AGN. The hosts of SN1999em and SN2012cw have AGNs as well, but the SNe are likely to be too distant for non-thermal contamination of the spectrum and as noted before SN2000ds is a Ca-rich SN. Unfortunately it is not possible to plot any of these SNe on the diagnostic BPT diagram  presented in Figure \ref{fig_bpt_diagram} since no H$\beta$ emission is detected. The effect of outliers on the statistical evaluation results of our N2 sample was checked by tentatively varying their metallicity. Any differences are negligible and thus we conclude that the validity ranges have no impact on our overall findings. 

\begin{table}
	\caption{Validity limits of M13 and PP04 calibrations as given in \protect\cite{marino2013} and \protect\cite{pettini2004}, respectively. N2 and O3N2 given by equations \ref{N2_indicator} and  \ref{O3N2_indicator}, respectively}
	\setlength{\tabcolsep}{0.1cm}
    \begin{tabular}{ccc}
        \hline
    	\hline
    	calibration & validity limits & 12+log(O/H) validity limits \\
    	\hline
          M13-N2 & -1.6 < N2 < -0.2 & 8.00 < 12+log(O/H) < 8.65 \\
       M13-O3N2 & -1.1 < O3N2 < 1.7 & 8.17 < 12+log(O/H) < 8.77 \\
          PP04-N2 & -2.5 < N2 < -0.3 & 7.48 < 12+log(O/H) < 8.73 \\
       PP04-O3N2 & -1.0 < O3N2 < 1.9 & 8.12 < 12+log(O/H) < 9.05 \\
    	\hline
    \end{tabular}%
    \label{tab_validity_limits}%
\end{table}%

To test the consistency between the two calibrations (M13-O3N2 and M13-N2) we compare the metallicity of each SN calculated with each calibration; the results are presented in Figure \ref{fig_scatterplot}. It is clear that there is generally a good linear agreement between the calibrations (linear regression coefficient $\sim$76\%) but an increasing scatter with increasing metallicity is present. This scatter may be caused by the relatively small number of targets in the lower metallicity range and/or by the limitations of the N2 calibration at higher metallicities.

\begin{figure*}
\includegraphics[trim={0.5cm 1cm 1cm 1.5cm}, clip, width=1.85\columnwidth, angle=0]{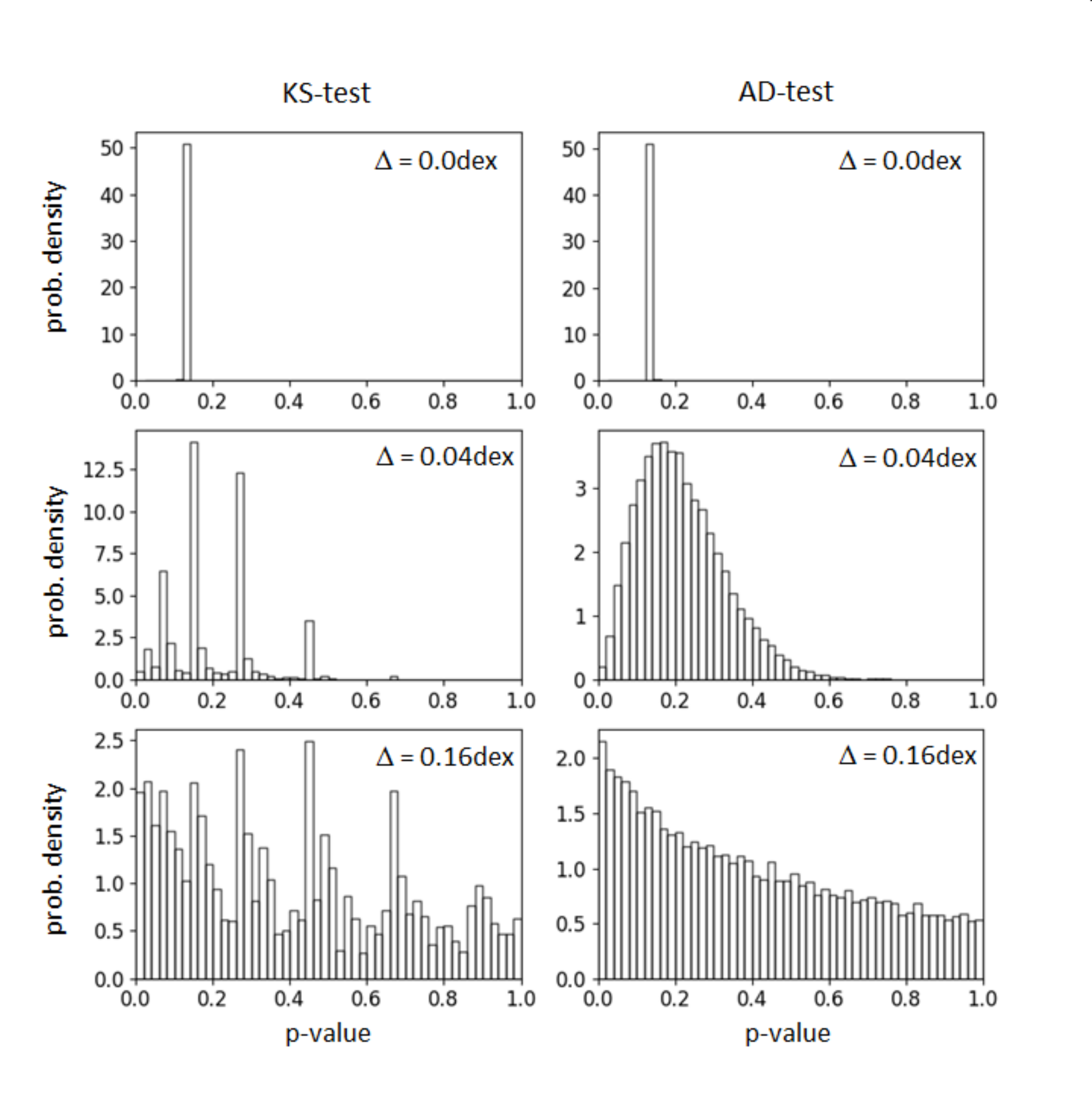}
\caption{Representative example for the results of the MC simulations, based on perturbing our sample with differing levels of uncertainty. The figure shows six results of the MC simulations with three results of the KS-test (left) and three results of the AD-test (right). In all cases, the same samples of metallicity values have been used (Type Ib and Type IIP M13-O3N2 results). The number N of runs was 20000 for all simulations. All plots show the probability density of p-values; the area of the histograms bins (bin width = 0.02) is normalized to one in all charts. The top row shows results for an uncertainty 0.0\,dex with a single bin for the p-value as expected. In the middle row with an uncertainty of 0.04\,dex (representative value for the observational and reduction process uncertainty) a significant broadening of the p-value distribution is visible already. Finally, in the bottom row with 0.16\,dex uncertainty (the M13-N2 calibration uncertainty) all p-values as outcome of the tests are possible with high probability. The result shown is qualitatively the same for all sample combinations of Table \ref{tab_m13testresults}.}
\label{fig_histogram}
\end{figure*}

The differences between the N2 and the O3N2 results may have a physical background. As highlighted by e.g. \cite{pettini2004} and \cite{marino2013}, the [N\,{\sc ii}] line saturates for metallicities greater than solar metallicity (12 + log(O/H) = 8.69, \citealt{asplund2009}) while the [O\,{\sc iii}] line does not. Consequently, the N2 calibration should be unreliable for high metallicities as it relies just on [N\,{\sc ii}]. This larger uncertainty of the N2 calibration at high metallicities is consistent with the larger scatter at higher metallicities in Figure \ref{fig_scatterplot}. However, the spread, potentially indicating saturation effects, starts at about 8.5\,dex, well below solar metallicity. The N2 calibration is known to produce scatter above solar metallicity but our work suggests that these effects could be present at lower metallicities of $\sim$8.5\,dex, however given the large calibration errors it is difficult to draw any firm conclusions. 

The number of outliers (6\% of total sample) is too small to explain the differences between the N2 and O3N2 results because the CDFs do not change significantly if the outliers are removed. The observed trends between SN subtypes in the CDFs are most likely not caused by an improper application of the strong emission lines methods.

\subsection{Monte-Carlo Simulations}    \label{sub_mcs}

Results of the statistical significance tests presented in Table \ref{tab_m13testresults} are based on values of the metallicities for environment of the three SN subtypes. These values suffer from large uncertainties of $\pm$0.16\,dex and $\pm$0.18\,dex for M13-N2 and M13-O3N2, respectively and can ultimately lead to a large range of absolute values and thus a range of p-values. For example, \citet{leloudas2011} performed Monte-Carlo (MC) simulations perturbing the metallicity around the mean value and found 0.007$<$p$<$0.483 at a 68\% confidence interval. Monte-Carlo (MC) simulations have been applied to estimate the sensitivity of the p-values of the KS- and AD-test to the uncertainties of the metallicity results from the INT data.

\begin{description}
\item{} The MC simulations were done by a simple algorithm as  follows: 
\item[$\bullet$] take the vector a\textsubscript{1} of the nominal metallicity values of the first SN type and add a random vector (representing the possible effect of the uncertainty), where each vector component is randomly drawn from a Gaussian distribution with mean=0.0 and standard deviation sigma=sd\textsubscript{1};
\item[$\bullet$] take the vector a\textsubscript{2} of the nominal metallicity values of the second SN type and add a random vector, where each vector component is randomly drawn from a Gaussian distribution with mean=0.0 and sigma=sd\textsubscript{2};
\item[$\bullet$] apply the KS- and the AD-test to the vector a\textsubscript{1} and a\textsubscript{2};
\item[$\bullet$] repeat the three steps above for a large number N;
\item[$\bullet$] create the histogram of the N calculated p-values of the KS- and of the AD-test.
\end{description}

The number N was chosen to be 20000 as a reasonable choice between significance of the MC results and the required time. The standard deviations sd\textsubscript{1} and sd\textsubscript{2} were constant for each particular MC simulation, but they were changed between 0.0\,dex and 0.18\,dex (maximum uncertainty of the calibrations) for different MC runs to evaluate the distribution of the p-values at different uncertainties. As in Section \ref{sect_res}, the ‘R’ package implementations of the KS- and the AD-test have been used to perform the tests in the MC runs.

The distributions of resulting p-values are presented in Figure \ref{fig_histogram} and the range of possible p-values increases as the uncertainty increases. For observational uncertainties of $\pm$0.04\,dex p-values up to $\sim$0.5 for the KS-test are possible, which is similar to the range found by \citet{leloudas2011}. For the AD-test, p-values range between 0 and 0.6. However, for the uncertainties given by the strong emission line calibrations ($\pm$0.16\,dex), all p-values between 0 and 1 are possible in both the KS- and AD-tests independent of the samples used. There is a clear qualitative difference between the distributions of p-values of the KS- and the AD-test: the KS-test distribution tends to a more multimodal distribution, the AD-test to a more continuous distribution of the p-values (see Figure \ref{fig_histogram} for an example). We suggest that this could be due to use of a single supremum value in the KS-test in combination with the relatively small samples. However, this is beyond the scope of this paper. Taking into account the high sensitivity of the p-values to uncertainties, the application of the KS-test and/or AD-test is questioned if the samples under test suffer large uncertainties. 

\subsection{Constraints for Type Ibc Progenitors}    \label{sub_con}

The results for mean metallicities in Table \ref{tab_statistics} show no statistically significant difference between the three SN subtypes which is in a good agreement with previous studies by \cite{anderson2010}, \cite{galbany2016c}, \cite{kuncarayakti2018}, \cite{leloudas2011} and \cite{sanders2012}. The results of \cite{modjaz2011} and \cite{kuncarayakti2013a}, which found higher metallicities for Type Ic SNe compared with Type Ib SNe, could not be confirmed. As discussed in Section \ref{sect_res}, there is only a slight metallicity difference of 0.08\,dex between the Type Ibc and Type IIP SN for the M13-O3N2 calibration. This tendency is confirmed in the corresponding CDF (Figure \ref{fig_o3n2_cdf}), but the KS- and the AD-test in Section \ref{sect_res} revealed no statistical significance between the parent population of different SN subtypes in agreement with the KS-test results of  \cite{anderson2010}, \cite{galbany2016c}, \cite{kuncarayakti2018}, \cite{leloudas2011}, \cite{sanders2012} and the AD-test results of \cite{galbany2018}.

However, the MC simulation results of KS- and AD-test highlight the limitations of these two statistical tests as a result of the large uncertainties associated with the calibrations themselves; this is consistent with the findings of \citet{leloudas2011} for the KS-test. Additionally, as for all statistical tests, if the test does not reject the null hypothesis, this does not conclusively mean that the alternative hypothesis is wrong.

\begin{figure*}
\begin{minipage}[t]{0.99\columnwidth}
\includegraphics[trim={0.3cm 0cm 1.35cm 0.7cm}, width=1.0\columnwidth]{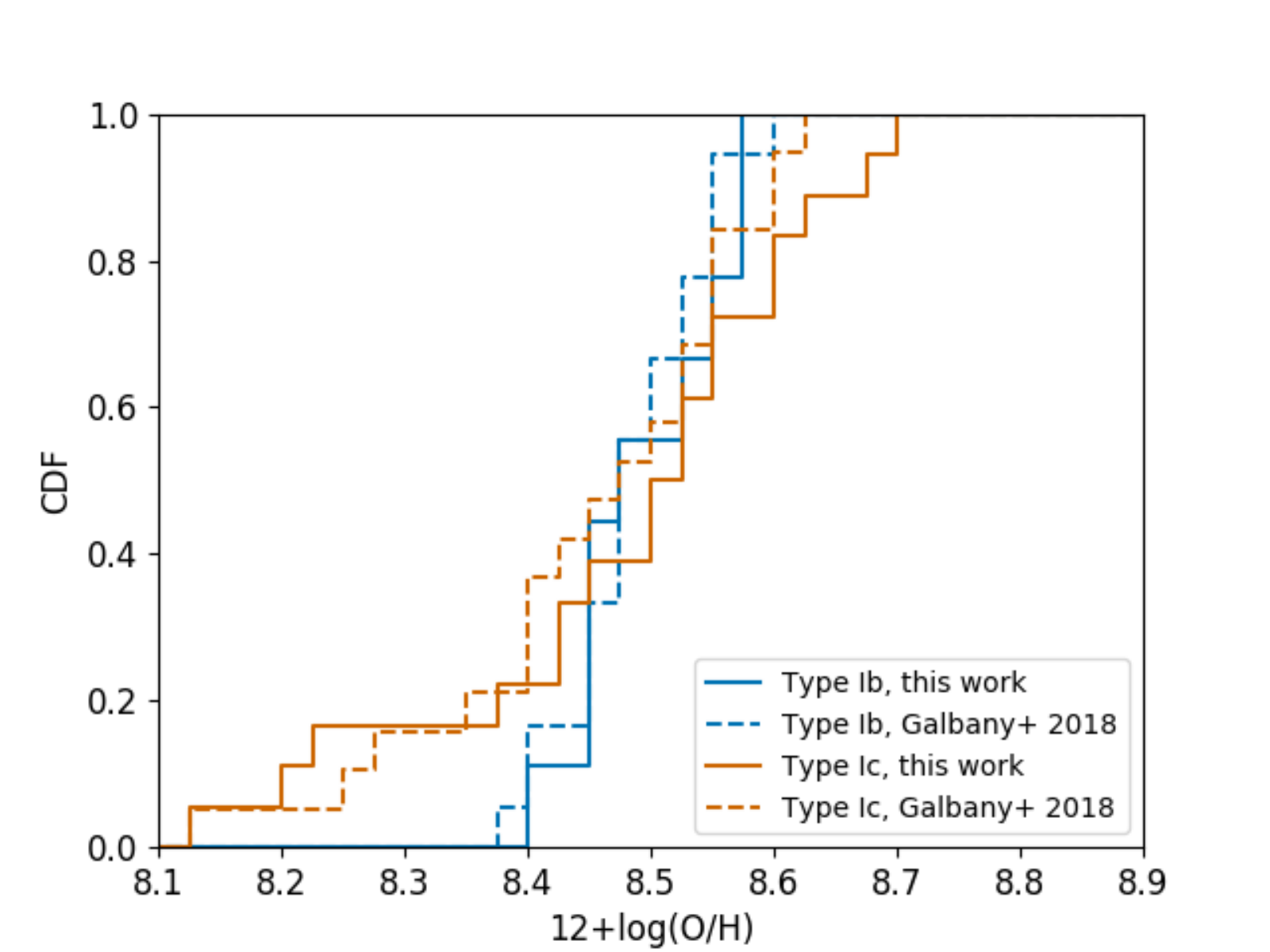}
\caption{Comparison of Type Ib and Type Ic with M13-O3N2 calibration CDFs from this work (solid lines) and from the Type Ibc sub-sample of \citet{galbany2018} (dashed lines). The CDFs for these almost completely distinct samples show very good agreement. Binning width for CDF calculation: 0.025\,dex.}
\label{fig_compared_cdfs}
\end{minipage}
\hspace{.09\columnwidth}
\begin{minipage}[t]{0.99\columnwidth}
\includegraphics[trim={0.3cm 0cm 1.35cm 0.7cm}, width=1.0\columnwidth]{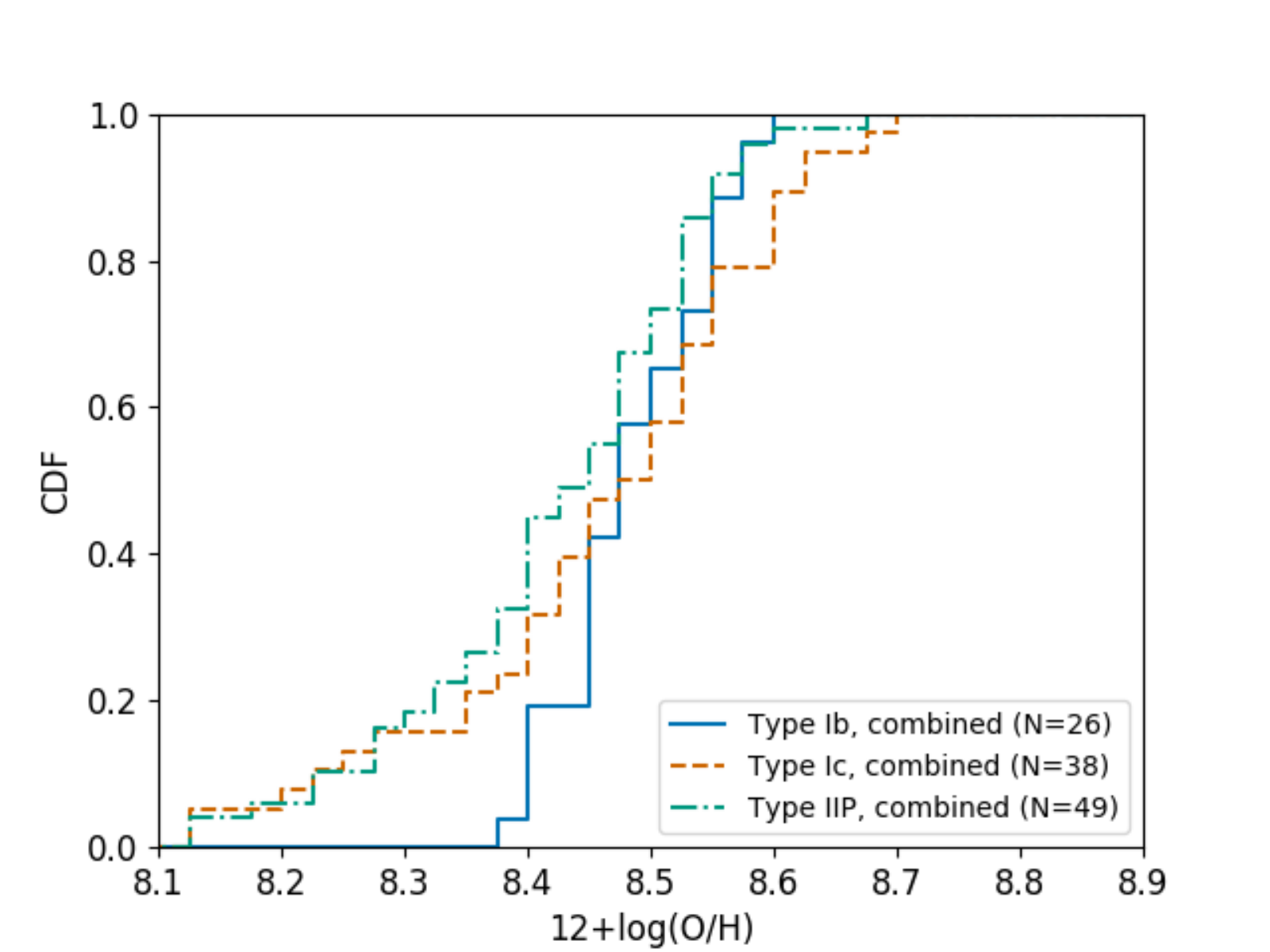}
\caption{M13-O3N2 calibration CDFs of the combined sample of this work (including targets of Table \ref{tab_opportunities}) and the Type Ib/Ic/IIP sub-sample of \citet{galbany2018}. The four targets common in the two M13-O3N2 samples are replaced by the average value of both metallicities. Binning width for CDF calculation: 0.025\,dex.}
\label{fig_combined_cdfs}
\end{minipage}
\end{figure*}

\begin{table}
\caption{Number of targets N, mean values and standard deviations $\sigma$ of the metallicities split into the three SN types and for the total sample based on the M13-N2 and M13-O3N2 calibrations for the results of our study combined with results from \citet{galbany2018}.}
    \begin{tabular}{c|ccc|ccc}
    \hline
    \hline
    SN    & \multicolumn{1}{c}{N(N2)} & \multicolumn{2}{c|}{M13-N2} & \multicolumn{1}{c}{N(O3N2)} & \multicolumn{2}{c}{M13-O3N2} \\
    type  &       & \multicolumn{2}{c|}{[12+log(O/H)]} &       & \multicolumn{2}{c}{[12+log(O/H)]} \\
    \multicolumn{1}{c|}{} &       & \multicolumn{1}{c} {mean} & \multicolumn{1}{c|}{$\sigma$} &       & \multicolumn{1}{c} {mean} & \multicolumn{1}{c}{$\sigma$} \\
    \hline
    Ib    & 31    & 8.52  & 0.06  & 26     & 8.50  & 0.06 \\
    Ic    & 40    & 8.49  & 0.13  & 38    & 8.47  & 0.14 \\
    IIP   & 60    & 8.49  & 0.13  & 49    & 8.44  & 0.12 \\
    \hline
    all   & 131    & 8.50  & 0.12  & 113    & 8.46  & 0.12 \\
    \hline
    \end{tabular}%
  \label{tab_statistics_comb}%
\end{table}%

For example, the distribution of Type Ib SNe metallicity environments is narrower than Type Ic and IIP (Figure \ref{fig_o3n2_cdf}). This finding is consistent with the CDFs of \cite{kuncarayakti2018} and \cite{galbany2018}  giving some evidence of different physical qualities of Type Ib environments which implies different physical progenitors of Type Ib SN compared to Type Ic and IIP. To test this further, results for the Type Ib/Ic/IIP sub-sample\footnote{the data of the total sample of \cite{galbany2018} have been downloaded from \url{https://iopscience.iop.org/article/10.3847/1538-4357/aaaf20}} in \citet{galbany2018} have been used. This sub-sample consists of 19 Type Ib, 20 Type Ic and 31 Type IIP, and it is almost completely distinct from our sample (only 5 IIP, 2 Ib and 1 Ic SNe are common in both samples). Figure \ref{fig_compared_cdfs} compares the Type Ib and Ic CDFs from this work with the corresponding CDFs of the Type Ib and Ic sub-sample of \cite{galbany2018} for the M13-O3N2 calibration. The narrow Type Ib distribution and the broader Type Ic distribution starting at low metallicities seen in our work are confirmed by the \cite{galbany2018} sub-sample. 

\begin{table*}
    \begin{threeparttable}
	\caption{P-values of the two-sample KS-test and two-sample AD-test for the M13 results of our study combined with results from \citet{galbany2018}. The results of statistical tests on the combined sample indicate a statistically significant difference of the O3N2 results of Type Ib vs. IIP and Type Ib vs. Ic+IIP for a significance level of 5\%. All other cases indicate no significant difference.}
  \setlength{\tabcolsep}{0.151cm}
    \begin{tabular}{c|cc|cc}
    	\hline
    	\hline
    	& \multicolumn{2}{c|}{KS-test} & \multicolumn{2}{c}{AD-test} \\
   		\hline
    	SN & M13-N2 & M13-O3N2 & M13-N2 & M13-O3N2 \\
 		types & p-value & p-value  & p-value & p-value \\
    	\hline
    	Ib-Ic & 0.245 & 0.311 & 0.138 & 0.086 \\
   		Ib-IIP & 0.120  & 0.027  & 0.077 & 0.020 \\
  		Ic-IIP & 0.788 & 0.189 &  0.853 & 0.165 \\
   		\hline
        Ib-(Ic+IIP) & 0.088 & 0.056  & 0.082 & 0.037 \\
        \hline
  \end{tabular}%
  \label{tab_m13testresults_comb}%
  \end{threeparttable}
\end{table*}%

Figure \ref{fig_combined_cdfs} shows the M13-O3N2 calibration CDFs of the combined sample from this work (including the 4 targets of Table \ref{tab_opportunities}) plus the \cite{galbany2018} sub-sample. The combined M13-O3N2 sample consists of 26 Type Ib, 38 Type Ic and 49 Type IIP. It confirms our CDFs of Figure \ref{fig_cdfs} both qualitatively and quantitatively with a higher statistical significance because of the considerably larger sample size. The mean metallicities and the standard deviations of the combined sample are shown in Table \ref{tab_statistics_comb}. The smaller standard deviation of the Type Ib results compared with Types Ic and IIP, as seen in Table \ref{tab_statistics}, is confirmed in the combined sample. The KS- and AD-test of the combined sample (Table \ref{tab_m13testresults_comb}) indicate a statistically significant difference of the O3N2 results between Type Ib vs. IIP and Type Ib vs. Ic+IIP at a significance level of 5\%. All other test results indicate no significant differences below a 5\% level. 

We note that these statistical tests are still subject to the large calibration uncertainties as discussed in Section \ref{sub_mcs}. However, accepting the limitations of the KS- and AD-tests, is it possible to draw any conclusions regarding the progenitors of the different SN subtypes based on the CDF's and subsequent analysis? If the differences are indeed real, what do they represent? 

Single massive stars have less physical parameters determining their mass loss (essentially initial mass, metallicity, rotation and magnetic field). Compared with single stars, compact binaries have additional orbital and evolutionary parameters of the companion influencing the mass-loss/overflow rate. For instance, a massive star with mass and metallicity to evolve as single star to a Type Ib SN, can evolve to Type Ic by additional mass-loss to its companion in a compact binary system - the number of evolutionary pathways of the massive star has been increased by the companion. Consequently, the larger number of physical parameters induces the expectation of a larger scatter (e.g. \citealt{xiao2019}) and consequently of broader CDFs in metallicity for the binary scenario. Consequently, the narrower Type Ib CDFs seen in the Figures \ref{fig_cdfs}, \ref{fig_compared_cdfs} and \ref{fig_combined_cdfs} may be an indication that the single massive progenitor star fraction for Type Ib is larger than for Type Ic. 

Type Ic SNe have a broad distribution starting at low metallicities, where it closely matches the Type IIP distribution for 12$+$log(O/H) $<$ 8.3. Besides the broad distribution, the low metallicity limit for Type Ic is a second indication of a significant binary fraction of Type Ic SNe progenitors. Stripping H- and He-shells at low metallicities by stellar winds only is unlikely because the mass loss strongly depends on metallicity (the higher metallicity the stronger mass loss; e.g. \citealt{vink2012}). Thus, the binary scenario is favoured as the mass stripping mechanism at low metallicities. This conclusion is consistent with \cite{smith2011b}, who derived, based on standard IMFs and the observed fractions of core-collapse SNe, the necessity of binary progenitors. However, the single massive star formation channel for Type Ic SNe cannot be ruled out from our data and a mixture of both formation channels is likely. 

While the agreement of the Type Ic and IIP CDFs at lower metallicity (e.g. this work; \citealt{galbany2018, kuncarayakti2018}) might seem contradictory given the single, massive nature of confirmed IIP progenitors (\citealt{smartt2009, vandyk2017}), Type IIP progenitors have no need to lose their outer envelope allowing a broader range of masses and metallicities for single massive progenitors of Type IIP SN compared with Type Ib.   

All constraints drawn above are not as distinctive for the N2 calibration results. However, it is reassuring that the CDFs created from data of the independent investigations by \cite{galbany2018} are very similar (Figure \ref{fig_compared_cdfs}) to ours and that the CDFs of the combined sample (Figure \ref{fig_combined_cdfs}) still show the distinctive features. The combined samples produce a statistically significant rejection of our null-hypothesis for Ib-IIP SNe with p\,=\,0.027 and p\,=\,0.020 (see Table \ref{tab_m13testresults_comb}) for the KS- and AD-tests, respectively, as a result of increased sample size. More observations of SNe environments and metallicity calibrations with significantly lower uncertainties, supported by direct measurements of metallicities via massive stars (\citealt{davies2017}) are still required to investigate all SN subtypes thoroughly and improve our understanding of core-collapse SNe progenitors.

\section{Summary and Conclusions}   \label{sect_con}

This work calculates and compares the metallicities of environments of Type IIP, Type Ib and Type Ic SNe within a luminosity distance of 30 Mpc by the strong emission line method. 76 targets were observed with INT/IDS and metallicity was measured for 65 of these using the N2 and O3N2 calibrations. The CDFs based on O3N2 calibration exhibit distinct features: narrow
Type Ib distribution, Type Ic at low metallicities and higher mean metallicities for Ibc SN compared to IIP. However, taking into account measurement errors, the statistical tests (KS-test and AD-test) are formally unable to reject the null hypothesis that the different SN types come from the same parent population. 

The results of our analysis have been confirmed and strengthened by the CDFs created from the Type Ib/Ic/IIP sample of \cite{galbany2018} as well as by the CDFs of the large combined sample of this work and the \cite{galbany2018} sub-sample. For the combined sample, both statistical tests indicate a statistically significant difference between the Type Ib and Type IIP M13-O3N2 results at 5\% significance level. This suggests that Ib and IIP SN do indeed have different progenitor populations. 
 
This statistical result combined with the apparent shapes of CDFs (see Figure \ref{fig_combined_cdfs}), might be explained by a significant fraction of single massive stars as Type Ib progenitors and suggests a significant fraction of compact binaries as progenitors of Type Ic SNe because they are present in low metallicity environments. Despite the large sample size of observed SNe environments these indications need further investigation with more observations of SNe environments in future work, to verify the significance of our results.

MC simulations have shown that the large calibration uncertainties have a significant effect on the range of p-values possible for both the KS- and AD-tests. These uncertainties must be reduced through improved calibrations and investigation of the saturation of the [N\,{\sc ii}] emission line.   

\section*{Acknowledgements}
The work presented is based on observations made with the Isaac Newton Telescope. The Isaac Newton Telescope is operated on the island of La Palma by the Isaac Newton Group of Telescopes in the Spanish Observatorio del Roque de los Muchachos of the Instituto de Astrofísica de Canarias. We would like to thank Isaac Newton Group of Telescopes staff and our co-observers A. Brocklebank, T. Davison, A. de Burgos, L. Holden, D. Nespral, S. Percival, T. Wilson, and T. Zegmott for their contributions to the observation data. J. L. Pledger, A. E. Sansom and S.M. Habergham-Mawson acknowledge financial support through the Panel for the Application of Telescope Time (PATT) travel grant (IDs ST/S005307/1 and ST/M00211X/1). Supernova data have been obtained from the Open Supernova Catalogue currently maintained by James Guillochon and Jerod Parrent. Host galaxy data have been obtained from the NASA/IPAC Extragalactic Database (NED), which is operated by the Jet Propulsion Laboratory, California Institute of Technology, under contract with the National Aeronautics and Space Administration. The research has made use of tools created by hard work of the astronomical community, acknowledged here by tool name and URL: IRAF (https://iraf.net), Starlink (http://starlink.eao.hawaii.edu/starlink), Hyperleda (http://leda.univ-lyon1.fr), 'R'-project (http://www.R-project.org), PPXF (https://pypi.org/project/ppxf), MILES stellar library (http://miles.iac.es/pages/stellar-libraries/miles-library.php)


\section*{Data Availability}
The data underlying this article will be shared on reasonable request to the corresponding author.



\bibliographystyle{mnras}
\bibliography{SNref}

\begin{thebibliography}{}
\makeatletter
\relax
\def\mn@urlcharsother{\let\do\@makeother \do\$\do\&\do\#\do\^\do\_\do\%\do\~}
\def\mn@doi{\begingroup\mn@urlcharsother \@ifnextchar [ {\mn@doi@}
  {\mn@doi@[]}}
\def\mn@doi@[#1]#2{\def\@tempa{#1}\ifx\@tempa\@empty \href
  {http://dx.doi.org/#2} {doi:#2}\else \href {http://dx.doi.org/#2} {#1}\fi
  \endgroup}
\def\mn@eprint#1#2{\mn@eprint@#1:#2::\@nil}
\def\mn@eprint@arXiv#1{\href {http://arxiv.org/abs/#1} {{\tt arXiv:#1}}}
\def\mn@eprint@dblp#1{\href {http://dblp.uni-trier.de/rec/bibtex/#1.xml}
  {dblp:#1}}
\def\mn@eprint@#1:#2:#3:#4\@nil{\def\@tempa {#1}\def\@tempb {#2}\def\@tempc
  {#3}\ifx \@tempc \@empty \let \@tempc \@tempb \let \@tempb \@tempa \fi \ifx
  \@tempb \@empty \def\@tempb {arXiv}\fi \@ifundefined
  {mn@eprint@\@tempb}{\@tempb:\@tempc}{\expandafter \expandafter \csname
  mn@eprint@\@tempb\endcsname \expandafter{\@tempc}}}

\bibitem[\protect\citeauthoryear{Aldering, Humphreys  \& Richmond}{Aldering
  et~al.}{1994}]{aldering1994}
Aldering G.,  Humphreys R.~M.,   Richmond M.,  1994, \mn@doi [AJ]
  {10.1086/116886}, 107, 662

\bibitem[\protect\citeauthoryear{Anderson \& James}{Anderson \&
  James}{2008}]{anderson2008}
Anderson J.~P.,  James P.~A.,  2008, \mn@doi [MNRAS]
  {10.1111/j.1365-2966.2008.13843.x}, 390, 1527

\bibitem[\protect\citeauthoryear{Anderson \& James}{Anderson \&
  James}{2009}]{anderson2009}
Anderson J.~P.,  James P.~A.,  2009, \mn@doi [MNRAS]
  {10.1111/j.1365-2966.2009.15324.x}, 399, 559

\bibitem[\protect\citeauthoryear{Anderson, Covarrubias, James, Hamuy  \&
  Habergham}{Anderson et~al.}{2010}]{anderson2010}
Anderson J.~P.,  Covarrubias R.~A.,  James P.~A.,  Hamuy M.,   Habergham S.~M.,
   2010, \mn@doi [MNRAS] {10.1111/j.1365-2966.2010.17118.x}, 407, 2660

\bibitem[\protect\citeauthoryear{Anderson et~al.,}{Anderson
  et~al.}{2014}]{anderson2014}
Anderson J.~P.,  et~al., 2014, \mn@doi [ApJ] {10.1088/0004-637X/786/1/67}, 786,
  67

\bibitem[\protect\citeauthoryear{Anderson, James, Habergham, Galbany  \&
  Kuncarayakti}{Anderson et~al.}{2015}]{anderson2015}
Anderson J.~P.,  James P.~A.,  Habergham S.~M.,  Galbany L.,   Kuncarayakti H.,
   2015, \mn@doi [PASA] {10.1017/pasa.2015.19}, 32, 19A

\bibitem[\protect\citeauthoryear{Arcavi et~al.,}{Arcavi
  et~al.}{2010}]{arcavi2010}
Arcavi I.,  et~al., 2010, \mn@doi [ApJ] {10.1088/0004-637X/721/1/777}, 721, 777

\bibitem[\protect\citeauthoryear{Asplund, Grevesse, Sauval  \& Scott}{Asplund
  et~al.}{2009}]{asplund2009}
Asplund M.,  Grevesse N.,  Sauval A.~J.,   Scott P.,  2009, \mn@doi [ARA\&A]
  {10.1146/annurev.astro.46.060407.145222}, 47, 481

\bibitem[\protect\citeauthoryear{Baldwin, Phillips  \& Terlevich}{Baldwin
  et~al.}{1981}]{baldwin1981}
Baldwin J.~A.,  Phillips M.~M.,   Terlevich R.,  1981, \mn@doi [PASP]
  {10.1086/130766}, 93, 5

\bibitem[\protect\citeauthoryear{Barth, van Dyk, Filippenko, Leibundgut  \&
  Richmond}{Barth et~al.}{1996}]{barth1996}
Barth A.~J.,  van Dyk S.~D.,  Filippenko A.~V.,  Leibundgut B.,   Richmond
  M.~W.,  1996, \mn@doi [AJ] {10.1086/117940}, 111, 2047

\bibitem[\protect\citeauthoryear{Benetti et~al.,}{Benetti
  et~al.}{2016}]{benetti2016}
Benetti S.,  et~al., 2016, \mn@doi [MNRAS] {10.1093/mnras/stv2811}, 456, 3296

\bibitem[\protect\citeauthoryear{Bersten et~al.,}{Bersten
  et~al.}{2014}]{bersten2014}
Bersten M.~C.,  et~al., 2014, \mn@doi [AJ] {10.1088/0004-6256/148/4/68}, 148,
  68

\bibitem[\protect\citeauthoryear{Boissier \& Prantzos}{Boissier \&
  Prantzos}{2009}]{boissier2009}
Boissier S.,  Prantzos N.,  2009, \mn@doi [A\&A] {10.1051/0004-6361/200811234},
  503, 137

\bibitem[\protect\citeauthoryear{Cao et~al.,}{Cao et~al.}{2013}]{cao2013}
Cao Y.,  et~al., 2013, \mn@doi [ApJ] {10.1088/2041-8205/775/1/L7}, 775, L7

\bibitem[\protect\citeauthoryear{Cappellari}{Cappellari}{2017}]{cappellari2017}
Cappellari M.,  2017, \mn@doi [MNRAS] {10.1093/mnras/stw3020}, 466, 798

\bibitem[\protect\citeauthoryear{Cappellari \& Emsellem}{Cappellari \&
  Emsellem}{2004}]{cappellari2004}
Cappellari M.,  Emsellem E.,  2004, \mn@doi [PASP] {10.1086/381875}, 116, 138

\bibitem[\protect\citeauthoryear{Chen}{Chen}{2021}]{chen2021}
Chen K.-J.,  2021, \mn@doi [IJMPD] {10.1142/S0218271821300019}, 30, 2130001

\bibitem[\protect\citeauthoryear{Crockett et~al.,}{Crockett
  et~al.}{2007}]{crockett2007}
Crockett R.~M.,  et~al., 2007, \mn@doi [MNRAS]
  {10.1111/j.1365-2966.2007.12283.x}, 381, 835

\bibitem[\protect\citeauthoryear{Crockett et~al.,}{Crockett
  et~al.}{2008}]{crockett2008}
Crockett R.~M.,  et~al., 2008, \mn@doi [MNRAS]
  {10.1111/j.1745-3933.2008.00540.x}, 391, L5

\bibitem[\protect\citeauthoryear{Crowther}{Crowther}{2007}]{crowther2007}
Crowther P.~A.,  2007, \mn@doi [ARA\&A]
  {10.1146/annurev.astro.45.051806.110615}, 45, 177

\bibitem[\protect\citeauthoryear{Crowther}{Crowther}{2013}]{crowther2013}
Crowther P.~A.,  2013, \mn@doi [MNRAS] {10.1093/mnras/sts145}, 428, 1927

\bibitem[\protect\citeauthoryear{Currie, Berry, Jenness, Gibb, Bell  \&
  Draper}{Currie et~al.}{2014}]{currie2014}
Currie M.~J.,  Berry D.~S.,  Jenness T.,  Gibb A.~G.,  Bell G.~S.,   Draper
  P.~W.,  2014, ASPC, 485, 391

\bibitem[\protect\citeauthoryear{Davies et~al.,}{Davies
  et~al.}{2017}]{davies2017}
Davies B.,  et~al., 2017, \mn@doi [ApJ] {10.3847/1538-4357/aa89ed}, 847, 112

\bibitem[\protect\citeauthoryear{Dessart, Hillier, Livne, Yoon, Woosley,
  Waldman  \& Langer}{Dessart et~al.}{2011}]{dessart2011}
Dessart L.,  Hillier D.~J.,  Livne E.,  Yoon S.-C.,  Woosley S.,  Waldman R.,
  Langer N.,  2011, \mn@doi [MNRAS] {10.1111/j.1365-2966.2011.18598.x}, 414,
  2985

\bibitem[\protect\citeauthoryear{Dowd}{Dowd}{2018}]{Rproject2018}
Dowd C.,  2018, twosamples: Fast Permutation Based Two Sample Tests.
\url {https://CRAN.R-project.org/package=twosamples}

\bibitem[\protect\citeauthoryear{Eldridge \& Maund}{Eldridge \&
  Maund}{2016}]{eldridge2016}
Eldridge J.~J.,  Maund J.~R.,  2016, \mn@doi [MNRAS] {10.1093/mnrasl/slw099},
  461, L117

\bibitem[\protect\citeauthoryear{Eldridge, Izzard  \& Tout}{Eldridge
  et~al.}{2008}]{eldridge2008}
Eldridge J.~J.,  Izzard R.~G.,   Tout C.~A.,  2008, \mn@doi [MNRAS]
  {10.1111/j.1365-2966.2007.12738.x}, 384, 1109

\bibitem[\protect\citeauthoryear{Eldridge, Fraser, Maund  \& Smartt}{Eldridge
  et~al.}{2015}]{eldridge2015}
Eldridge J.~J.,  Fraser M.,  Maund J.~R.,   Smartt S.~J.,  2015, \mn@doi
  [MNRAS] {10.1093/mnras/stu2197}, 446, 2689

\bibitem[\protect\citeauthoryear{{Elias-Rosa} et~al.,}{{Elias-Rosa}
  et~al.}{2010}]{elisarosa2010}
{Elias-Rosa} N.,  et~al., 2010, \mn@doi [\apjl] {10.1088/2041-8205/714/2/L254},
  \href {https://ui.adsabs.harvard.edu/abs/2010ApJ...714L.254E} {714, L254}

\bibitem[\protect\citeauthoryear{Elias-Rosa et~al.,}{Elias-Rosa
  et~al.}{2011}]{eliasrosa2011}
Elias-Rosa N.,  et~al., 2011, \mn@doi [AJ] {10.1088/0004-637X/742/1/6}, 742, 6

\bibitem[\protect\citeauthoryear{Elias-Rosa et~al.,}{Elias-Rosa
  et~al.}{2013}]{eliasrosa2013}
Elias-Rosa N.,  et~al., 2013, \mn@doi [MNRAS] {10.1093/mnrasl/slt124}, 436,
  L109

\bibitem[\protect\citeauthoryear{Fang \& Maeda}{Fang \& Maeda}{2018}]{fang2018}
Fang Q.,  Maeda K.,  2018, \mn@doi [ApJ] {10.3847/1538-4357/aad096}, 864, 47

\bibitem[\protect\citeauthoryear{Filippenko}{Filippenko}{1997}]{filippenko1997}
Filippenko A.~V.,  1997, ARA\&A, 35, 309

\bibitem[\protect\citeauthoryear{Filippenko, Chornock, Swift, Modjaz, Simcoe
  \& Rauch}{Filippenko et~al.}{2003}]{filippenko2003}
Filippenko A.~V.,  Chornock R.,  Swift B.,  Modjaz M.,  Simcoe R.,   Rauch M.,
  2003, IAUC, 8159, 2

\bibitem[\protect\citeauthoryear{{Fitzpatrick}}{{Fitzpatrick}}{1985}]{Fitzpatrick1985}
{Fitzpatrick} E.~L.,  1985, \mn@doi [\apj] {10.1086/163694}, \href
  {https://ui.adsabs.harvard.edu/abs/1985ApJ...299..219F} {299, 219}

\bibitem[\protect\citeauthoryear{Folatelli et~al.,}{Folatelli
  et~al.}{2016}]{folatelli2016}
Folatelli G.,  et~al., 2016, \mn@doi [ApJL] {10.3847/2041-8205/825/2/L22}, 825,
  L22

\bibitem[\protect\citeauthoryear{Foley, Berger, Fox, Levesque, Challis, Ivans,
  Rhoads  \& Soderberg}{Foley et~al.}{2011}]{foley2011}
Foley R.~J.,  Berger E.,  Fox O.,  Levesque E.~M.,  Challis P.~J.,  Ivans
  I.~I.,  Rhoads J.~E.,   Soderberg A.~M.,  2011, \mn@doi [AJ]
  {10.1088/0004-637X/732/1/32}, 732, 32

\bibitem[\protect\citeauthoryear{Fremling et~al.,}{Fremling
  et~al.}{2014}]{fremling2014}
Fremling C.,  et~al., 2014, \mn@doi [A\&A] {10.1051/0004-6361/201423884}, 565,
  A114

\bibitem[\protect\citeauthoryear{Fruchter et~al.,}{Fruchter
  et~al.}{2006}]{fruchter2006}
Fruchter A.~S.,  et~al., 2006, \mn@doi [Nature] {10.1038/nature04787}, 441, 463

\bibitem[\protect\citeauthoryear{Gal-Yam}{Gal-Yam}{2019}]{galyam2019}
Gal-Yam A.,  2019, \mn@doi [ARA\&A] {10.1146/annurev-astro-081817-051819}, 57,
  305

\bibitem[\protect\citeauthoryear{Gal-Yam \& Leonard}{Gal-Yam \&
  Leonard}{2009}]{galyam2009}
Gal-Yam A.,  Leonard D.~C.,  2009, \mn@doi [Nature] {10.1038/nature07934}, 458,
  865

\bibitem[\protect\citeauthoryear{Gal-Yam et~al.,}{Gal-Yam
  et~al.}{2005}]{galyam2005}
Gal-Yam A.,  et~al., 2005, \mn@doi [ApJ] {10.1086/491622}, 630, L29

\bibitem[\protect\citeauthoryear{Galbany et~al.,}{Galbany
  et~al.}{2016a}]{galbany2016a}
Galbany L.,  et~al., 2016a, \mn@doi [AJ] {10.3847/0004-6256/151/2/33}, 151, 33

\bibitem[\protect\citeauthoryear{Galbany et~al.,}{Galbany
  et~al.}{2016b}]{galbany2016b}
Galbany L.,  et~al., 2016b, \mn@doi [MNRAS] {10.1093/mnras/stv2620}, 455, 4087

\bibitem[\protect\citeauthoryear{Galbany et~al.,}{Galbany
  et~al.}{2016c}]{galbany2016c}
Galbany L.,  et~al., 2016c, \mn@doi [A\&A] {10.1051/0004-6361/201528045}, 591,
  A48

\bibitem[\protect\citeauthoryear{Galbany et~al.,}{Galbany
  et~al.}{2018}]{galbany2018}
Galbany L.,  et~al., 2018, \mn@doi [ApJ] {10.3847/1538-4357/aaaf20}, 855, 107

\bibitem[\protect\citeauthoryear{{Graham}}{{Graham}}{2019}]{Graham2019}
{Graham} J.~F.,  2019, arXiv e-prints, \href
  {https://ui.adsabs.harvard.edu/abs/2019arXiv190513197G} {p. arXiv:1905.13197}

\bibitem[\protect\citeauthoryear{Groh, Georgy  \& Ekstrom}{Groh
  et~al.}{2013}]{groh2013}
Groh J.~H.,  Georgy C.,   Ekstrom S.,  2013, \mn@doi [A\&A]
  {10.1051/0004-6361/201322369}, 558, L1

\bibitem[\protect\citeauthoryear{Guillochon, Parrent, Kelley  \&
  Margutti}{Guillochon et~al.}{2017}]{guillochon2017}
Guillochon J.,  Parrent J.,  Kelley L.~Z.,   Margutti R.,  2017, \mn@doi [ApJ]
  {10.3847/1538-4357/835/1/64}, 835, 64

\bibitem[\protect\citeauthoryear{Gutiérrez et~al.,}{Gutiérrez
  et~al.}{2017}]{gutierrez2017}
Gutiérrez C.~P.,  et~al., 2017, \mn@doi [ApJ] {10.3847/1538-4357/aa8f52}, 850,
  89

\bibitem[\protect\citeauthoryear{Hakobyan et~al.,}{Hakobyan
  et~al.}{2014}]{hakobyan2014}
Hakobyan A.~A.,  et~al., 2014, \mn@doi [MNRAS] {10.1093/mnras/stu1598}, 444,
  2428

\bibitem[\protect\citeauthoryear{Hakobyan et~al.,}{Hakobyan
  et~al.}{2016}]{hakobyan2016}
Hakobyan A.~A.,  et~al., 2016, \mn@doi [MNRAS] {10.1093/mnras/stv2853}, 456,
  2848

\bibitem[\protect\citeauthoryear{Helou, Madore, Schmitz, Bicay, Wu  \&
  Bennett}{Helou et~al.}{1991}]{helou1991}
Helou G.,  Madore B.~F.,  Schmitz M.,  Bicay M.~D.,  Wu X.,   Bennett J.,
  1991, \mn@doi [ASSL] {10.1007/978-94-011-3250-3_10}, 171, 89

\bibitem[\protect\citeauthoryear{Henry \& Worthey}{Henry \&
  Worthey}{1999}]{henry1999}
Henry R. B.~C.,  Worthey G.,  1999, \mn@doi [PASP] {10.1086/316403}, 111, 919

\bibitem[\protect\citeauthoryear{Hirai}{Hirai}{2017}]{hirai2017}
Hirai R.,  2017, \mn@doi [MNRAS] {10.1093/mnras/stw3321}, 466, 3775

\bibitem[\protect\citeauthoryear{{Hummer} \& {Storey}}{{Hummer} \&
  {Storey}}{1987}]{Hummer1987}
{Hummer} D.~G.,  {Storey} P.~J.,  1987, \mn@doi [\mnras]
  {10.1093/mnras/224.3.801}, \href
  {https://ui.adsabs.harvard.edu/abs/1987MNRAS.224..801H} {224, 801}

\bibitem[\protect\citeauthoryear{James \& Anderson}{James \&
  Anderson}{2006}]{james2006}
James P.~A.,  Anderson J.~P.,  2006, \mn@doi [A\&A]
  {10.1051/0004-6361:20054509}, 453, 57

\bibitem[\protect\citeauthoryear{Johnson, Kochanek  \& Adams}{Johnson
  et~al.}{2017}]{johnson2017}
Johnson S.~A.,  Kochanek C.~S.,   Adams S.~M.,  2017, \mn@doi [MNRAS]
  {10.1093/mnras/stx2170}, 472, 3115

\bibitem[\protect\citeauthoryear{Kelly \& Kirshner}{Kelly \&
  Kirshner}{2012}]{kelly2012}
Kelly P.~L.,  Kirshner R.~P.,  2012, \mn@doi [ApJ]
  {10.1088/0004-637X/759/2/107}, 759, 107

\bibitem[\protect\citeauthoryear{Kewley \& Ellison}{Kewley \&
  Ellison}{2008}]{kewley2008}
Kewley L.~J.,  Ellison S.~L.,  2008, \mn@doi [ApJ] {10.1086/587500}, 681, 1183

\bibitem[\protect\citeauthoryear{Kewley, Dopita, Sutherland, Heisler  \&
  Trevena}{Kewley et~al.}{2001}]{Kewley2001}
Kewley L.~J.,  Dopita M.~A.,  Sutherland R.~S.,  Heisler C.~A.,   Trevena J.,
  2001, \mn@doi [ApJ] {10.1086/321545}, 556, 121

\bibitem[\protect\citeauthoryear{Kewley, Nicholls  \& Sutherland}{Kewley
  et~al.}{2019}]{kewley2019}
Kewley L.~J.,  Nicholls D.~C.,   Sutherland R.~S.,  2019, \mn@doi [ARA\&A]
  {10.1146/annurev-astro-081817-051832}, 57, 511

\bibitem[\protect\citeauthoryear{Kilpatrick et~al.,}{Kilpatrick
  et~al.}{2018}]{kilpatrick2018}
Kilpatrick C.~D.,  et~al., 2018, \mn@doi [MNRAS] {10.1093/mnras/sty2022}, 480,
  2072

\bibitem[\protect\citeauthoryear{Kilpatrick et~al.,}{Kilpatrick
  et~al.}{2021}]{kilpatrick2021}
Kilpatrick C.~D.,  et~al., 2021, \mn@doi [MNRAS] {10.1093/mnras/stab838}, 504,
  2073

\bibitem[\protect\citeauthoryear{Krühler, Kuncarayakti, Schady, Anderson,
  Galbany  \& Gensior}{Krühler et~al.}{2017}]{kruhler2017}
Krühler T.,  Kuncarayakti H.,  Schady P.,  Anderson J.~P.,  Galbany L.,
  Gensior J.,  2017, \mn@doi [A\&A] {10.1051/0004-6361/201630268}, 602, A85

\bibitem[\protect\citeauthoryear{Kudritzki \& Puls}{Kudritzki \&
  Puls}{2000}]{kudritzki2000}
Kudritzki R.-P.,  Puls J.,  2000, ARA\&A, 38, 613

\bibitem[\protect\citeauthoryear{Kuncarayakti et~al.,}{Kuncarayakti
  et~al.}{2013a}]{kuncarayakti2013b}
Kuncarayakti H.,  et~al., 2013a, \mn@doi [AJ] {10.1088/0004-6256/146/2/31},
  146, 31

\bibitem[\protect\citeauthoryear{Kuncarayakti et~al.,}{Kuncarayakti
  et~al.}{2013b}]{kuncarayakti2013a}
Kuncarayakti H.,  et~al., 2013b, \mn@doi [AJ] {10.1088/0004-6256/146/2/30},
  146, 30

\bibitem[\protect\citeauthoryear{Kuncarayakti et~al.,}{Kuncarayakti
  et~al.}{2015}]{kuncarayakti2015}
Kuncarayakti H.,  et~al., 2015, \mn@doi [A\&A] {10.1051/0004-6361/201425604},
  579, A95

\bibitem[\protect\citeauthoryear{Kuncarayakti et~al.,}{Kuncarayakti
  et~al.}{2018}]{kuncarayakti2018}
Kuncarayakti H.,  et~al., 2018, \mn@doi [A\&A] {10.1051/0004-6361/201731923},
  613, A35

\bibitem[\protect\citeauthoryear{Leloudas et~al.,}{Leloudas
  et~al.}{2011}]{leloudas2011}
Leloudas G.,  et~al., 2011, \mn@doi [A\&A] {10.1051/0004-6361/201116692}, 530,
  A95

\bibitem[\protect\citeauthoryear{Lunnan et~al.,}{Lunnan
  et~al.}{2017}]{lunnan2017}
Lunnan R.,  et~al., 2017, \mn@doi [ApJ] {10.3847/1538-4357/836/1/60}, 836, 60

\bibitem[\protect\citeauthoryear{Lyman, James, Perets, Anderson, Gal-Yam,
  Mazzali  \& Percival}{Lyman et~al.}{2013}]{lyman2013}
Lyman J.~D.,  James P.~A.,  Perets H.~B.,  Anderson J.~P.,  Gal-Yam A.,
  Mazzali P.,   Percival S.~M.,  2013, \mn@doi [MNRAS] {10.1093/mnras/stt1038},
  434, 527

\bibitem[\protect\citeauthoryear{Makarov, Prugniel, Terekhova, Courtois  \&
  Vauglin}{Makarov et~al.}{2014}]{makarov2014}
Makarov D.,  Prugniel P.,  Terekhova N.,  Courtois H.,   Vauglin I.,  2014,
  \mn@doi [A\&A] {10.1051/0004-6361/201423496}, 570, A13

\bibitem[\protect\citeauthoryear{Maoz, Mannucci  \& Nelemans}{Maoz
  et~al.}{2014}]{maoz2014}
Maoz D.,  Mannucci F.,   Nelemans G.,  2014, \mn@doi [ARA\&A]
  {10.1146/annurev-astro-082812-141031}, 52, 107

\bibitem[\protect\citeauthoryear{Marino et~al.,}{Marino
  et~al.}{2013}]{marino2013}
Marino R.~A.,  et~al., 2013, \mn@doi [A\&A] {10.1051/0004-6361/201321956}, 559,
  A114

\bibitem[\protect\citeauthoryear{Marsaglia, Tsang  \& Wang}{Marsaglia
  et~al.}{2003}]{marsaglia2003}
Marsaglia G.,  Tsang W.~W.,   Wang J.,  2003, \mn@doi [JSS]
  {10.18637/jss.v008.i18}, 8, 1

\bibitem[\protect\citeauthoryear{Massey}{Massey}{1952}]{massey1952}
Massey F.~J.,  1952, \mn@doi [AoMS] {10.1214/aoms/1177729388}, 23, 435

\bibitem[\protect\citeauthoryear{Maund}{Maund}{2018}]{maund2018}
Maund J.~R.,  2018, \mn@doi [MNRAS] {10.1093/mnras/sty093}, 476, 2629

\bibitem[\protect\citeauthoryear{Maund \& Smartt}{Maund \&
  Smartt}{2005}]{maund2005a}
Maund J.~R.,  Smartt S.~J.,  2005, \mn@doi [MNRAS]
  {10.1111/j.1365-2966.2005.09034.x}, 360, 288

\bibitem[\protect\citeauthoryear{Maund, Smartt, Kudritzki, Podsiadlowski  \&
  Gilmore}{Maund et~al.}{2004}]{maund2004}
Maund J.~R.,  Smartt S.~J.,  Kudritzki R.~P.,  Podsiadlowski P.,   Gilmore
  G.~F.,  2004, \mn@doi [Nature] {10.1038/nature02161}, 427, 129

\bibitem[\protect\citeauthoryear{Maund, Smartt  \& Schweizer}{Maund
  et~al.}{2005}]{maund2005b}
Maund J.~R.,  Smartt S.~J.,   Schweizer F.,  2005, \mn@doi [ApJ]
  {10.1086/491620}, 630, L33

\bibitem[\protect\citeauthoryear{Maund et~al.,}{Maund et~al.}{2011}]{maund2011}
Maund J.~R.,  et~al., 2011, \mn@doi [ApJ] {10.1088/2041-8205/739/2/L37}, 739,
  L37

\bibitem[\protect\citeauthoryear{Mazzali et~al.,}{Mazzali
  et~al.}{2002}]{mazzali2002}
Mazzali P.~A.,  et~al., 2002, \mn@doi [ApJ] {10.1086/341504}, 572, L61

\bibitem[\protect\citeauthoryear{Modjaz et~al.,}{Modjaz
  et~al.}{2008}]{modjaz2008}
Modjaz M.,  et~al., 2008, \mn@doi [AJ] {10.1088/0004-6256/135/4/1136}, 135,
  1136

\bibitem[\protect\citeauthoryear{Modjaz, Kewley, Bloom, Filippenko, Perley  \&
  Silverman}{Modjaz et~al.}{2011}]{modjaz2011}
Modjaz M.,  Kewley L.,  Bloom J.~S.,  Filippenko A.~V.,  Perley D.,   Silverman
  J.~M.,  2011, \mn@doi [ApJ] {10.1088/2041-8205/731/1/L4}, 731, L4

\bibitem[\protect\citeauthoryear{Osterbrock \& Ferland}{Osterbrock \&
  Ferland}{2006}]{osterbrock2006}
Osterbrock D.~E.,  Ferland G.~J.,  2006, Astrophysics of gaseous nebulae and
  active galactic nuclei, 2nd ed edn.
University Science Books, Sausalito, Calif

\bibitem[\protect\citeauthoryear{Pandey et~al.,}{Pandey
  et~al.}{2021}]{pandey2021}
Pandey S.~B.,  et~al., 2021, arXiv:astro-ph/2106.15856

\bibitem[\protect\citeauthoryear{Pettini \& Pagel}{Pettini \&
  Pagel}{2004}]{pettini2004}
Pettini M.,  Pagel B. E.~J.,  2004, \mn@doi [MNRAS]
  {10.1111/j.1365-2966.2004.07591.x}, 348, L59

\bibitem[\protect\citeauthoryear{Pettitt}{Pettitt}{1976}]{pettitt1976}
Pettitt A.~N.,  1976, \mn@doi [Biometrika] {10.2307/2335097}, 63, 161

\bibitem[\protect\citeauthoryear{Podsiadlowski, Joss  \& Hsu}{Podsiadlowski
  et~al.}{1992}]{podsiadlowski1992}
Podsiadlowski P.,  Joss P.~C.,   Hsu J. J.~L.,  1992, \mn@doi [ApJ]
  {10.1086/171341}, 391, 246

\bibitem[\protect\citeauthoryear{Prantzos \& Boissier}{Prantzos \&
  Boissier}{2003}]{prantzos2003}
Prantzos N.,  Boissier S.,  2003, \mn@doi [A\&A] {10.1051/0004-6361:20030717},
  406, 259

\bibitem[\protect\citeauthoryear{Press}{Press}{1988}]{press1988}
Press W.~H.,  ed. 1988, Numerical recipes in {C}: the art of scientific
  computing.
Cambridge University Press, Cambridge [Cambridgeshire] ; New York

\bibitem[\protect\citeauthoryear{Prieto, Stanek  \& Beacom}{Prieto
  et~al.}{2008}]{prieto2008}
Prieto J.~L.,  Stanek K.~Z.,   Beacom J.~F.,  2008, \mn@doi [ApJ]
  {10.1086/524654}, 673, 999

\bibitem[\protect\citeauthoryear{Sanders et~al.,}{Sanders
  et~al.}{2012}]{sanders2012}
Sanders N.~E.,  et~al., 2012, \mn@doi [ApJ] {10.1088/0004-637X/758/2/132}, 758,
  132

\bibitem[\protect\citeauthoryear{Sanders et~al.,}{Sanders
  et~al.}{2015}]{sanders2015}
Sanders N.~E.,  et~al., 2015, \mn@doi [ApJ] {10.1088/0004-637X/799/2/208}, 799,
  208

\bibitem[\protect\citeauthoryear{Schady, Eldridge, Anderson, Chen, Galbany,
  Kuncarayakti  \& Xiao}{Schady et~al.}{2019}]{schady2019}
Schady P.,  Eldridge J.~J.,  Anderson J.,  Chen T.-W.,  Galbany L.,
  Kuncarayakti H.,   Xiao L.,  2019, \mn@doi [MNRAS] {10.1093/mnras/stz2843},
  490, 4515

\bibitem[\protect\citeauthoryear{Schlafly \& Finkbeiner}{Schlafly \&
  Finkbeiner}{2011}]{schlafly2011}
Schlafly E.~F.,  Finkbeiner D.~P.,  2011, \mn@doi [ApJ]
  {10.1088/0004-637X/737/2/103}, 737, 103

\bibitem[\protect\citeauthoryear{Schlegel}{Schlegel}{1990}]{schlegel1990}
Schlegel E.~M.,  1990, MNRAS, 244, 269

\bibitem[\protect\citeauthoryear{Shen, Quataert  \& Pakmor}{Shen
  et~al.}{2019}]{shen2019}
Shen K.~J.,  Quataert E.,   Pakmor R.,  2019, \mn@doi [ApJ]
  {10.3847/1538-4357/ab5370}, 887, 180

\bibitem[\protect\citeauthoryear{Smartt}{Smartt}{2009}]{smartt2009}
Smartt S.~J.,  2009, \mn@doi [ARA\&A] {10.1146/annurev-astro-082708-101737},
  47, 63

\bibitem[\protect\citeauthoryear{Smartt}{Smartt}{2015}]{smartt2015}
Smartt S.~J.,  2015, \mn@doi [PASA] {10.1017/pasa.2015.17}, 32, e016

\bibitem[\protect\citeauthoryear{Smartt, Vreeswijk, Ramirez-Ruiz, Gilmore,
  Meikle, Ferguson  \& Knapen}{Smartt et~al.}{2002}]{smartt2002}
Smartt S.~J.,  Vreeswijk P.~M.,  Ramirez-Ruiz E.,  Gilmore G.~F.,  Meikle W.
  P.~S.,  Ferguson A. M.~N.,   Knapen J.~H.,  2002, \mn@doi [ApJ]
  {10.1086/341747}, 572, L147

\bibitem[\protect\citeauthoryear{Smith}{Smith}{2014}]{smith2014}
Smith N.,  2014, \mn@doi [ARA\&A] {10.1146/annurev-astro-081913-040025}, 52,
  487

\bibitem[\protect\citeauthoryear{Smith, Li, Filippenko  \& Chornock}{Smith
  et~al.}{2011a}]{smith2011b}
Smith N.,  Li W.,  Filippenko A.~V.,   Chornock R.,  2011a, \mn@doi [MNRAS]
  {10.1111/j.1365-2966.2011.17229.x}, 412, 1522

\bibitem[\protect\citeauthoryear{Smith et~al.,}{Smith
  et~al.}{2011b}]{smith2011a}
Smith N.,  et~al., 2011b, \mn@doi [AJ] {10.1088/0004-637X/732/2/63}, 732, 63

\bibitem[\protect\citeauthoryear{Taddia et~al.,}{Taddia
  et~al.}{2013a}]{taddia2013a}
Taddia F.,  et~al., 2013a, \mn@doi [A\&A] {10.1051/0004-6361/201321180}, 555,
  A10

\bibitem[\protect\citeauthoryear{Taddia et~al.,}{Taddia
  et~al.}{2013b}]{taddia2013b}
Taddia F.,  et~al., 2013b, \mn@doi [A\&A] {10.1051/0004-6361/201322276}, 558,
  A143

\bibitem[\protect\citeauthoryear{Taddia et~al.,}{Taddia
  et~al.}{2015}]{taddia2015}
Taddia F.,  et~al., 2015, \mn@doi [A\&A] {10.1051/0004-6361/201525989}, 580,
  A131

\bibitem[\protect\citeauthoryear{Taddia et~al.,}{Taddia
  et~al.}{2016}]{taddia2016}
Taddia F.,  et~al., 2016, \mn@doi [A\&A] {10.1051/0004-6361/201527983}, 587, L7

\bibitem[\protect\citeauthoryear{Taddia et~al.,}{Taddia
  et~al.}{2019}]{taddia2019}
Taddia F.,  et~al., 2019, \mn@doi [A\&A] {10.1051/0004-6361/201834429}, 621,
  A71

\bibitem[\protect\citeauthoryear{Tartaglia et~al.,}{Tartaglia
  et~al.}{2017}]{tartaglia2017}
Tartaglia L.,  et~al., 2017, \mn@doi [ApJ] {10.3847/2041-8213/aa5c7f}, 836, L12

\bibitem[\protect\citeauthoryear{Teffs, Prentice, Mazzali  \& Ashall}{Teffs
  et~al.}{2021}]{teffs2021}
Teffs J.~J.,  Prentice S.~J.,  Mazzali P.~A.,   Ashall C.,  2021, \mn@doi
  [MNRAS] {10.1093/mnras/stab258}, 502, 3829

\bibitem[\protect\citeauthoryear{Tody}{Tody}{1986}]{tody1986}
Tody D.,  1986, \mn@doi [SPIE] {10.1117/12.968154}, 627, 733

\bibitem[\protect\citeauthoryear{Turatto}{Turatto}{2003}]{turatto2003}
Turatto M.,  2003, \mn@doi [arXiv:astro-ph/0301107] {10.1007/3-540-45863-8_3}

\bibitem[\protect\citeauthoryear{Valenti et~al.,}{Valenti
  et~al.}{2011}]{valenti2011}
Valenti S.,  et~al., 2011, \mn@doi [MNRAS] {10.1111/j.1365-2966.2011.19262.x},
  416, 3138

\bibitem[\protect\citeauthoryear{Valenti et~al.,}{Valenti
  et~al.}{2016}]{valenti2016}
Valenti S.,  et~al., 2016, \mn@doi [MNRAS] {10.1093/mnras/stw870}, 459, 3939

\bibitem[\protect\citeauthoryear{Van~Dyk}{Van~Dyk}{2017}]{vandyk2017}
Van~Dyk S.~D.,  2017, \mn@doi [RSPTA] {10.1098/rsta.2016.0277}, 375, 20160277

\bibitem[\protect\citeauthoryear{Van~Dyk \& Schuyler}{Van~Dyk \&
  Schuyler}{1992}]{vandyk1992}
Van~Dyk S.~D.,  Schuyler D.,  1992, \mn@doi [AJ] {10.1086/116195}, 103, 1788

\bibitem[\protect\citeauthoryear{Van~Dyk, Garnavich, Filippenko, Höflich,
  Kirshner, Kurucz  \& Challis}{Van~Dyk et~al.}{2002}]{vandyk2002}
Van~Dyk S.~D.,  Garnavich P.~M.,  Filippenko A.~V.,  Höflich P.,  Kirshner
  R.~P.,  Kurucz R.~L.,   Challis P.,  2002, \mn@doi [PASP] {10.1086/344382},
  114, 1322

\bibitem[\protect\citeauthoryear{Van~Dyk et~al.,}{Van~Dyk
  et~al.}{2011}]{vandyk2011}
Van~Dyk S.~D.,  et~al., 2011, \mn@doi [ApJ] {10.1088/2041-8205/741/2/L28}, 741,
  L28

\bibitem[\protect\citeauthoryear{Van~Dyk et~al.,}{Van~Dyk
  et~al.}{2014}]{vandyk2014}
Van~Dyk S.~D.,  et~al., 2014, \mn@doi [AJ] {10.1088/0004-6256/147/2/37}, 147,
  37

\bibitem[\protect\citeauthoryear{Van~Dyk, de Mink  \& Zapartas}{Van~Dyk
  et~al.}{2016}]{vandyk2016}
Van~Dyk S.~D.,  de Mink S.~E.,   Zapartas E.,  2016, \mn@doi [ApJ]
  {10.3847/0004-637X/818/1/75}, 818, 75

\bibitem[\protect\citeauthoryear{Van~Dyk et~al.,}{Van~Dyk
  et~al.}{2018}]{vandyk2018}
Van~Dyk S.~D.,  et~al., 2018, \mn@doi [ApJ] {10.3847/1538-4357/aac32c}, 860, 90

\bibitem[\protect\citeauthoryear{Vazdekis, Sánchez-Blázquez, Falcón-Barroso,
  Cenarro, Beasley, Cardiel, Gorgas  \& Peletier}{Vazdekis
  et~al.}{2010}]{vazdekis2010}
Vazdekis A.,  Sánchez-Blázquez P.,  Falcón-Barroso J.,  Cenarro A.~J.,
  Beasley M.~A.,  Cardiel N.,  Gorgas J.,   Peletier R.~F.,  2010, \mn@doi
  [MNRAS] {10.1111/j.1365-2966.2010.16407.x}, 404, 1639

\bibitem[\protect\citeauthoryear{Vink \& Gräfener}{Vink \&
  Gräfener}{2012}]{vink2012}
Vink J.~S.,  Gräfener G.,  2012, \mn@doi [ApJ] {10.1088/2041-8205/751/2/L34},
  751, L34

\bibitem[\protect\citeauthoryear{Vink \& Sander}{Vink \&
  Sander}{2021}]{vink2021}
Vink J.~S.,  Sander A. A.~C.,  2021, \mn@doi [MNRAS] {10.1093/mnras/stab902},
  504, 2051

\bibitem[\protect\citeauthoryear{Vink \& de Koter}{Vink \&
  de~Koter}{2005}]{vink2005}
Vink J.~S.,  de Koter A.,  2005, \mn@doi [A\&A] {10.1051/0004-6361:20052862},
  442, 587

\bibitem[\protect\citeauthoryear{Woosley \& Bloom}{Woosley \&
  Bloom}{2006}]{woosley2006}
Woosley S.~E.,  Bloom J.~S.,  2006, ARA\&A, 44, 507

\bibitem[\protect\citeauthoryear{Xiang et~al.,}{Xiang et~al.}{2019}]{xiang2019}
Xiang D.,  et~al., 2019, \mn@doi [ApJ] {10.3847/1538-4357/aaf8b0}, 871, 176

\bibitem[\protect\citeauthoryear{Xiao, Galbany, Eldridge  \& Stanway}{Xiao
  et~al.}{2019}]{xiao2019}
Xiao L.,  Galbany L.,  Eldridge J.~J.,   Stanway E.~R.,  2019, \mn@doi [MNRAS]
  {10.1093/mnras/sty2557}, 482, 384

\bibitem[\protect\citeauthoryear{Yoon, Woosley  \& Langer}{Yoon
  et~al.}{2010}]{yoon2010}
Yoon S.-C.,  Woosley S.~E.,   Langer N.,  2010, \mn@doi [ApJ]
  {10.1088/0004-637X/725/1/940}, 725, 940

\makeatother
\end{thebibliography}

\onecolumn
\section*{Appendix}

\setcounter{table}{0}
\renewcommand{\thetable}{A.\arabic{table}}
\begin{longtable}{@{\extracolsep{\fill}}cccccccccc@{}}
\caption{List of observed targets. The columns are the target name, target SN type, host galaxy name and morphology,  luminosity distance D\textsubscript{L}, absolute magnitude M\textsubscript{V} of the host galaxies, the position angle PA of the target with respect to galaxy centre, the distance d\textsubscript{c} of the target from host centre, the number N of observations and the total exposure time.} \\
\hline \hline
 target & type  & host  & host type & D\textsubscript{L}    &  M\textsubscript{V}    &  PA   &d\textsubscript{C} & N  & Exp. time\\
          &       &   &    &  [Mpc]     & [mag] & [\textdegree] &  [\arcsec]     &       & [sec]    \\
\hline
\endfirsthead
\caption[]{(continued)}\\
\hline \hline
    target & type  & host  & host type & D\textsubscript{L}    &  M\textsubscript{V}    &  PA   &d\textsubscript{C} & N  & Exp. time\\
          &       &    &   &  [Mpc]     & [mag] & [\textdegree] &  [\arcsec]     &       & [sec]    \\
\hline
\endhead
\tabularnewline
    1995bb & Ib/c  & PGC1409128 & Irr?  & 25.8  & -     & 87.6  & 9.38  & 1     & 4800 \\
    1998bv & IIP  & HS1035+4758 & ?     & 23.4  & -16.0 & 215.8 & 5.18  & 1     & 3600 \\
    1998dl & IIP  & NGC1084 & SA(s)c & 16.7  & -21.8 & 54.1  & 28.82 & 1     & 3600 \\
    1999eh & Ib    & NGC2770 & SA(s)c & 26.0  & -21.7 & 238.0 & 15.27 & 1     & 3600 \\
    1999em & IIP  & NGC1637 & SAB(rs)c & 7.7   & -20.6 & 227.0 & 23.76 & 1     & 3600 \\
    1999ev & IIP  & NGC4274 & SB(r)ab & 21.0  & -21.7 & 315.6 & 42.29 & 1     & 3600 \\
    1999gi & IIP  & NGC3184 & SAB(rs)cd & 11.0  & -21.0 & 358.5 & 61.22 & 1     & 4000 \\
    2000ds & Ib    & NGC2768 & E6    & 17.0  & -21.7 & 196.5 & 33.17 & 2     & 7200 \\
    2000ew & Ic    & NGC3810 & SA(rs)c & 16.0  & -21.4 & 187.5 & 20.27 & 1     & 3600 \\
    2001B & Ib    & IC391 & SA(s)c & 25.0  & -20.4 & 238.5 & 6.70  & 3     & 10800 \\
    2001ci & Ic    & NGC3079 & SB(s)c & 15.0  & -22.1 & 351.8 & 27.88 & 2     & 7200 \\
    2001fv & IIP  & NGC3512 & SAB(rs)c & 25.0  & -20.5 & 223.5 & 23.85 & 1     & 4200 \\
    2002hh & IIP  & NGC6946 & SAB(rs)cd & 4.7   & -14.8 & 207.5 & 129.63 & 1     & 3600 \\
    2002ji & Ib    & NGC3655 & SA(s)c & 26.0  & -21.7 & 236.1 & 25.11 & 1     & 3600 \\
    2002jz & Ic    & UGC2984 & SBdm  & 22.9  & -     & 197.5 & 3.88  & 2     & 7200 \\
    2003gd & IIP  & NGC628 & SA(s)c & 3.4   & -20.4 & 175.6 & 159.57 & 1     & 3600 \\
    2003ie & IIP  & NGC4051 & SAB(rs)bc & 13.0  & -21.3 & 101.4 & 93.26 & 1     & 1200 \\
    2003J & IIP  & NGC4157 & SAB(s)b & 15.7  & -22.0 & 242.5 & 71.89 & 1     & 5400 \\
    2003Z & IIP  & NGC2742 & SA(s)c & 20.0  & -21.6 & 346.3 & 32.42 & 1     & 5400 \\
    2004A & IIP  & NGC6207 & SA(s)c & 20.0  & -20.5 & 305.7 & 26.59 & 1     & 3600 \\
    2004ao & Ib    & UGC10862 & SB(rs)c & 26.0  & -19.1 & 165.3 & 26.36 & 2     & 7200 \\
    2004bm & Ic    & NGC3437 & SAB(rs)c & 21.0  & -21.3 & 296.1 & 5.69  & 1     & 3600 \\
    2004C & Ic    & NGC3683 & SB(s)c & 35.0  & -20.8 & 303.3 & 20.40 & 1     & 3600 \\
    2004dg & IIP  & NGC5806 & SAB(s)b & 22.0  & -22.0 & 261.0 & 21.85 & 1     & 3600 \\
    2004dj & IIP  & NGC2403 & SAB(s)cd & 3.5   & -19.4 & 94.0  & 159.14 & 1     & 3600 \\
    2004dk & Ib    & NGC6118 & SA(s)cd & 20.0  & -21.8 & 6.6   & 42.99 & 1     & 3600 \\
    2004ez & IIP  & NGC3430 & SAB(rs)c & 26.0  & -21.7 & 68.8  & 50.08 & 1     & 4600 \\
    2004fc & IIP  & NGC701 & SB(rs)c & 19.0  & -20.9 & 19.4  & 2.23  & 1     & 3600 \\
    2004gk & Ic    & IC3311 & Sdm   & 17.0  & -18.3 & 31.6  & 3.64  & 1     & 3600 \\
    2004gn & Ic    & NGC4527 & SAB(s)bc & 12.6  & -20.7 & 68.9  & 59.43 & 2     & 7200 \\
    2004gq & Ib    & NGC1832 & SB(r)bc & 16.0  & -21.7 & 45.0  & 30.83 & 1     & 3600 \\
    2004gt & Ic    & NGC4038 & SB(s)m & 16.0  & -21.8 & 255.4 & 38.57 & 1     & 3600 \\
    2005ad & IIP  & NGC941 & SAB(rs)c & 20.0  & -19.1 & 28.3  & 52.23 & 1     & 3600 \\
    2005ay & IIP  & NGC3938 & SA(s)c & 18.0  & -20.7 & 194.2 & 58.38 & 1     & 3600 \\
    2005cs & IIP  & NGC5194 & SA(s)bc & 6.1   & -20.9 & 179.3 & 67.30 & 1     & 3600 \\
    2005cz & Ib    & NGC4589 & E2    & 20.0  & -21.6 & 119.2 & 13.34 & 1     & 3800 \\
    2005kl & Ic    & NGC4369 & SA(rs)a & 21.0  & -21.1 & 306.7 & 7.52  & 1     & 4800 \\
    2005V & Ib/c  & NGC2146 & SB(s)ab & 17.0  & -22.1 & 24.8  & 4.19  & 1     & 3600 \\
    2006bp & IIP  & NGC3953 & SB(r)bc & 17.0  & -22.3 & 33.9  & 112.49 & 1     & 4000 \\
    2007aa & IIP  & NGC4030 & SA(s)bc & 23.0  & -22.3 & 41.9  & 91.88 & 1     & 3600 \\
    2007av & IIP  & NGC3279 & Sd    & 29.0  & -21.8 & 155.0 & 12.90 & 2     & 7800 \\
    2007C & Ib    & NGC4981 & SAB(r)bc & 21.0  & -21.5 & 157.9 & 23.75 & 1     & 4200 \\
    2007gr & Ic    & NGC1058 & SA(rs)c & 10.0  & -17.5 & 303.1 & 28.75 & 3     & 10800 \\
    2007od & IIP  & UGC12846 & Sm    & 24.0  & -     & 132.7 & 51.90 & 1     & 3600 \\
    2008D & Ib    & NGC2770 & SA(s)c & 26.0  & -21.7 & 325.3 & 67.27 & 2     & 7800 \\
    2008X & IIP  & NGC4141 & SBcd  & 28.0  & -18.9 & 60.1  & 9.22  & 1     & 3600 \\
    2009em & Ic    & NGC157 & SAB(rs)bc & 23.0  & -21.7 & 251.8 & 33.90 & 1     & 3600 \\
    2009js & IIP  & NGC918 & SAB(rs)c & 16.0  & -21.4 & 240.6 & 41.15 & 1     & 3600 \\
    2010br & Ib/c  & NGC4051 & SAB(rs)bc & 13.0  & -21.3 & 124.4 & 17.50 & 1     & 4000 \\
    2010io & Ic    & UGC4543 & SAdm  & 29.0  & -18.5 & 333.0 & 8.98  & 3     & 8400 \\
    2011ck & IIP  & NGC5425 & Sd    & 30.7  & -19.2 & 297.0 & 16.30 & 1     & 3600 \\
    2011dq & IIP  & NGC337 & SB(s)d & 24.4  & -21.1 & 300.4 & 40.51 & 1     & 3600 \\
    2011jm & Ic    & NGC4809 & Im    & 13.8  & -17.3 & 93.8  & 1.50  & 1     & 3600 \\
    2012A & IIP  & NGC3239 & IB(s)m & 8.8   & -19.7 & 133.9 & 49.57 & 1     & 4200 \\
    2012au & Ib    & NGC4790 & SB(rs)c & 20.0  & -19.4 & 66.5  & 4.51  & 1     & 4200 \\
    2012bv & IIP  & NGC6796 & Sbc   & 32.5  & -21.0 & 187.4 & 30.25 & 1     & 4800 \\
    2012cw & Ic    & NGC3166 & SAB(rs)a & 19.9  & -22.6 & 46.3  & 48.65 & 2     & 8400 \\
    2012ec & IIP  & NGC1084 & SA(s)c & 16.7  & -21.8 & 358.9 & 15.00 & 1     & 3600 \\
    2012fh & Ic    & NGC3344 & SAB(r)bc & 8.6   & -20.4 & 160.1 & 118.03 & 1     & 4200 \\
    2012P & Ib/c  & NGC5806 & SAB(s)b & 22.0  & -22.0 & 256.5 & 19.73 & 1     & 4000 \\
    2013ab & IIP  & NGC5669 & SAB(rs)d & 23.6  & -20.5 & 133.1 & 20.04 & 1     & 3600 \\
    2013bu & IIP  & NGC7331 & SA(s)bc & 12.1  & -23.0 & 204.0 & 55.62 & 2     & 8400 \\
    2013dk & Ic    & NGC4038 & SB(s)m & 16.0  & -21.8 & 194.6 & 15.81 & 1     & 4200 \\
    2013ej & IIP  & NGC628 & SA(s)c & 3.4   & -20.4 & 134.1 & 127.87 & 1     & 4800 \\
    2013ff & Ic    & NGC2748 & SAbc  & 21.9  & -21.0 & 215.6 & 24.84 & 1     & 3600 \\
    2013ge & Ic    & NGC3287 & SB(s)d & 19.3  & -18.4 & 18.7  & 50.56 & 2     & 6600 \\
    2014A & IIP  & NGC5054 & SA(s)bc & 27.0  & -21.4 & 57.1  & 14.72 & 1     & 3600 \\
    2014bc & IIP  & NGC4258 & SAB(s)bc & 6.6   & -21.9 & 141.7 & 3.44  & 1     & 4200 \\
    2014bi & IIP  & NGC4096 & SAB(rs)c & 8.3   & -21.1 & 20.4  & 54.95 & 1     & 4200 \\
    2014C & Ib    & NGC7331 & SA(s)bc & 12.1  & -23.0 & 140.6 & 31.19 & 1     & 3600 \\
    2014cx & IIP  & NGC337 & SB(s)d & 24.4  & -21.1 & 303.6 & 40.51 & 1     & 3600 \\
    2015aq & IIP  & UGC5015 & SABdm & 24.0  & -18.5 & 267.5 & 43.05 & 1     & 4200 \\
    2015V & IIP  & UGC11000 & S?    & 20.3  & -18.5 & 142.1 & 8.87  & 1     & 3600 \\
    2016bau & Ib    & NGC3631 & SA(s)c & 17.1  & -19.5 & 293.5 & 37.95 & 1     & 3600 \\
    2017ein & Ic    & NGC3938 & SA(s)c & 18.0  & -20.7 & 74.2  & 42.53 & 1     & 3600 \\
    2017iro & Ib    & NGC5480 & SA(s)c & 27.5  & -21.3 & 125.0 & 17.74 & 1     & 3600 \\
\hline
\label{tab_observations}
\end{longtable}

\newpage

\setcounter{footnote}{0}
\begin{longtable}{@{\extracolsep{\fill}}cccccccccc@{}}
\caption{Metallicities of the observed SNe environments. The columns are the target name, target SN type, the J2000.0 target coordinates and the measured environment metallicities by N2 and O3N2 method based on the M13 calibration. The metallicities based on the older PP04 calibration are given as well for better comparability with previous studies. The uncertainties are dominated by the calibration uncertainties in all cases, which are $\pm$0.16 dex for M13-N2, $\pm$0.18 dex for M13-O3N2, $\pm$0.18 dex for PP04-N2 and $\pm$0.14 dex for PP04-O3N2 (all uncertainties are 1$\sigma$ values as given in \citealt{marino2013} and \citealt{pettini2004}, respectively and correspond to the uncertainty on the final metallicity).} \\
\hline \hline
    target & type  & SN RA   & SN Dec & M13-N2 & M13-O3N2 & PP04-N2 & PP04-O3N2 \\
          &       & [h m s] & [\textdegree\,\arcmin\,\arcsec] & 12+log(O/H) & 12+log(O/H) & 12+log(O/H) & 12+log(O/H) \\
\hline
\endfirsthead
\caption[]{(continued)}\\
\hline \hline
    target & type  & SN RA   & SN Dec & M13-N2 & M13-O3N2 & PP04-N2 & PP04-O3N2 \\
          &       & [h m s] & [\textdegree\,\arcmin\,\arcsec] & 12+log(O/H) & 12+log(O/H) & 12+log(O/H) & 12+log(O/H) \\
\hline
\endhead
\tabularnewline
    1998bv & IIP  & 10 38 25.40 & +47 42 32.8 & 8.19  & 8.19  & 8.22  & 8.21 \\
    1998dl & IIP  & 02 46 01.47 & -07 34 25.1 & 8.52  & 8.52  & 8.62  & 8.72 \\
    1999eh & Ib    & 09 09 32.67 & +33 07 16.9 & 8.50  & -     & 8.61  & - \\
    1999em$^{\star}$ & IIP  & 04 41 27.04 & -02 51 45.2 & 8.77  & -     & 8.93  & - \\
    1999gi & IIP  & 10 18 16.66 & +41 26 28.2 & 8.42  & -     & 8.51  & - \\
    2000ds\footnote{classified as Ca-rich and excluded from statistical evalutions (see Section \ref{sect_res})}$^{\star}$ & Ib    & 09 11 36.24 & +60 01 42.2 & 8.85  & -     & 9.03  & - \\
    2000ew & Ic    & 11 40 58.52 & +11 27 55.9 & 8.50  & 8.71  & 8.60  & 8.99 \\
    2001B & Ib    & 04 57 19.24 & +78 11 16.5 & 8.51  & 8.49  & 8.61  & 8.65 \\
    2001ci & Ic    & 10 01 57.33 & +55 41 14.6 & 8.59  & 8.57  & 8.71  & 8.79 \\
    2001fv & IIP  & 11 04 01.66 & +28 01 55.7 & 8.71  & -     & 8.86  &  \\
    2002hh & IIP  & 20 34 44.29 & +60 07 19.0 & 8.52  & 8.54  & 8.63  & 8.74 \\
    2002ji & Ib    & 11 22 53.15 & +16 35 10.0 & 8.53  & 8.54  & 8.64  & 8.74 \\
    2002jz & Ic    & 04 13 12.52 & +13 25 07.3 & 8.25  & 8.21  & 8.29  & 8.24 \\
    2003gd & IIP  & 01 36 42.65 & +15 44 20.9 & 8.68  & -     & 8.83  & - \\
    2003J & IIP  & 12 10 57.72 & +50 28 31.8 & 8.67  & -     & 8.81  & - \\
    2003Z & IIP  & 09 07 32.46 & +60 29 17.5 & 8.55  & 8.40  & 8.66  & 8.53 \\
    2004A & IIP  & 16 43 01.90 & +36 50 12.5 & 8.50  & -     & 8.59  & - \\
    2004ao & Ib    & 17 28 09.35 & +07 24 55.5 & 8.49  & 8.41  & 8.59  & 8.55 \\
    2004bm & Ic    & 10 52 35.33 & +22 56 05.5 & 8.56  & 8.69  & 8.68  & 8.97 \\
    2004C & Ic    & 11 27 29.72 & +56 52 48.2 & 8.52  & 8.52  & 8.63  & 8.72 \\
    2004dg & IIP  & 14 59 58.96 & +01 53 25.6 & 8.53  & 8.68  & 8.64  & 8.96 \\
    2004dj & IIP  & 07 37 17.02 & +65 35 57.8 & 8.50  & 8.36  & 8.61  & 8.47 \\
    2004dk & Ib    & 16 21 48.93 & -02 16 17.3 & 8.37  & 8.46  & 8.44  & 8.63 \\
    2004fc & IIP  & 01 51 03.85 & -09 42 06.9 & 8.55  & 8.56  & 8.66  & 8.78 \\
    2004gk & Ic    & 12 25 33.23 & +12 15 40.1 & 8.50  & 8.44  & 8.60  & 8.59 \\
    2004gn & Ic    & 12 34 12.10 & +02 39 34.4 & 8.51  & 8.64  & 8.62  & 8.89 \\
    2004gq & Ib    & 05 12 04.81 & -15 40 54.2 & 8.55  & 8.58  & 8.66  & 8.80 \\
    2004gt & Ic    & 12 01 50.37 & -18 52 12.7 & 8.53  & 8.52  & 8.63  & 8.71 \\
    2005ay & IIP  & 11 52 48.07 & +44 06 18.4 & 8.56  & 8.59  & 8.68  & 8.81 \\
    2005cs & IIP  & 13 29 52.78 & +47 10 35.7 & 8.44  & -     & 8.52  & - \\
    2005kl & Ic    & 12 24 35.68 & +39 23 03.5 & 8.51  & 8.53  & 8.61  & 8.73 \\
    2005V & Ib/c  & 06 18 38.28 & +78 21 28.8 & 8.59  & 8.56  & 8.71  & 8.76 \\
    2006bp & IIP  & 11 53 55.74 & +52 21 09.4 & 8.54  & 8.54  & 8.65  & 8.75 \\
    2007aa & IIP  & 12 00 27.69 & -01 04 51.6 & 8.50  & 8.50  & 8.61  & 8.68 \\
    2007av & IIP  & 10 34 43.17 & +11 11 38.3 & 8.59  & -     & 8.72  & - \\
    2007C & Ib    & 13 08 49.30 & -06 47 01.0 & 8.59  & 8.58  & 8.71  & 8.80 \\
    2007gr & Ic    & 02 43 27.98 & +37 20 44.7 & 8.50  & 8.56  & 8.60  & 8.77 \\
    2008D & Ib    & 09 09 30.65 & +33 08 20.3 & 8.46  & 8.45  & 8.55  & 8.61 \\
    2008X & IIP  & 12 09 48.33 & +58 51 01.6 & 8.28  & 8.24  & 8.33  & 8.29 \\
    2009em & Ic    & 00 34 44.53 & -08 23 57.6 & 8.52  & 8.53  & 8.63  & 8.73 \\
    2009js & IIP  & 02 25 48.28 & +18 29 25.8 & 8.49  & -     & 8.59  & - \\
    2010io & Ic    & 08 43 21.41 & +45 44 18.0 & 8.31  & 8.24  & 8.36  & 8.29 \\
    2011ck & IIP  & 14 00 46.24 & +48 26 45.4 & 8.46  & 8.43  & 8.55  & 8.57 \\
    2011dq & IIP  & 00 59 47.75 & -07 34 20.5 & 8.36  & 8.30  & 8.42  & 8.38 \\
    2011jm & Ic    & 12 54 51.10 & +02 39 14.9 & 8.15  & 8.14  & 8.17  & 8.14 \\
    2012A & IIP  & 10 25 07.39 & +17 09 14.6 & 8.20  & 8.15  & 8.23  & 8.16 \\
    2012au & Ib    & 12 54 52.18 & -10 14 50.2 & 8.50  & 8.46  & 8.60  & 8.62 \\
    2012bv & IIP  & 19 21 30.36 & +61 08 12.0 & 8.58  & 8.48  & 8.70  & 8.66 \\
    2012cw$^{\star}$ & Ic    & 10 13 47.95 & +03 26 02.6 & 8.83  & -     & 9.00  & - \\
    2012ec & IIP  & 02 45 59.88 & -07 34 27.0 & 8.52  & 8.50  & 8.63  & 8.67 \\
    2012fh & Ic    & 10 43 34.05 & +24 53 29.0 & 8.47  & 8.43  & 8.57  & 8.57 \\
    2012P & Ib/c  & 14 59 59.12 & +01 53 24.4 & 8.50  & -     & 8.60  & - \\
    2013ab & IIP  & 14 32 44.49 & +09 53 12.3 & 8.57  & -     & 8.69  & - \\
    2013bu & IIP  & 22 37 02.17 & +34 24 05.2 & 8.53  & 8.49  & 8.64  & 8.67 \\
    2013dk & Ic    & 12 01 52.72 & -18 52 18.3 & 8.58  & 8.61  & 8.70  & 8.85 \\
    2013ej & IIP  & 01 36 48.16 & +15 45 31.0 & 8.66  & -     & 8.80  & - \\
    2013ff & Ic    & 09 13 38.88 & +76 28 10.8 & 8.51  & 8.47  & 8.61  & 8.64 \\
    2013ge & Ic    & 10 34 48.46 & +21 39 41.9 & 8.45  & 8.38  & 8.54  & 8.50 \\
    2014A & IIP  & 13 16 59.36 & -16 37 57.0 & 8.55  & -     & 8.67  & - \\
    2014bc$^{\star}$ & IIP  & 12 18 57.71 & +47 18 11.3 & 8.78  & -     & 8.95  & - \\
    2014bi & IIP  & 12 06 02.99 & +47 29 33.5 & 8.60  & -     & 8.72  & - \\
    2014C & Ib    & 22 37 05.60 & +34 24 31.9 & 8.65  & -     & 8.79  & - \\
    2014cx & IIP  & 00 59 47.83 & -07 34 18.6 & 8.36  & 8.28  & 8.43  & 8.35 \\
    2015V & IIP  & 17 49 27.05 & +36 08 36.0 & 8.36  & 8.29  & 8.42  & 8.37 \\
    2016bau & Ib    & 11 20 59.00 & +53 10 25.6 & 8.54  & -     & 8.65  & - \\
    2017ein & Ic    & 11 52 53.25 & +44 07 26.2 & 8.52  & 8.61  & 8.63  & 8.84 \\
\hline
\multicolumn{8}{l}{$^{\star}$ These SN metallicities lie outside of the validity range of the M13-N2 and/or PP04-N2 calibrations. See } \\   
\multicolumn{8}{l}{\,\,\,\,\,\,Section \ref{sub_rel} for details.}
\label{tab_results}
\end{longtable}

\begin{longtable}{@{\extracolsep{\fill}}cccccccccc@{}}
\caption{The metallicities of the four SNe environments observed by opportunity. See caption of Table \ref{tab_results} for the description of the columns and the uncertainties. These targets are not part of the statistical evaluation of our sample as they do not fulfill the selection criteria of the project, but they are included in the combined sample of our study and the Type Ib/Ic/IIP sample of \citet{galbany2018}.} \\
\hline \hline
    target & type  & SN RA   & SN Dec & M13-N2 & M13-O3N2 & PP04-N2 & PP04-O3N2 \\
          &       & [h m s] & [\textdegree\,\arcmin\,\arcsec] & 12+log(O/H) & 12+log(O/H) & 12+log(O/H) & 12+log(O/H) \\
\hline
\endfirsthead
\tabularnewline
    1990aa & Ic    & 00 52 59.22 & +29 01 48.3 & 8.40  & 8.37  & 8.47  & 8.49 \\
    1991ar & Ib    & 00 43 56.71 & +01 51 13.5 & 8.49  & 8.42  & 8.59  & 8.56 \\
    1996D & Ic    & 04 34 00.30 & -08 34 44.0 & 8.54  & 8.45  & 8.65  & 8.61 \\
    2009ga & IIP  & 23 28 26.78 & +22 24 50.4 & 8.54  & 8.54  & 8.65  & 8.74 \\
\hline
\label{tab_opportunities}%
\end{longtable}%






\bsp	
\label{lastpage}
\end{document}